\documentclass[reprint,superscriptaddress,amsmath,amssymb,aps,pra]{revtex4-2}
\usepackage{graphicx}
\usepackage{dcolumn}
\usepackage{bm}
\usepackage{tikz}
\usepackage{quantikz}
\usepackage{pgfplots}
\usepackage{booktabs}
\usepgfplotslibrary{colorbrewer,colormaps,external,fillbetween}
\pgfplotsset{
     compat=newest,
     colormap={inferno}{
     rgb(0)=(0.001462, 0.000466, 0.013866),
     rgb(15)=(0.037668, 0.025921, 0.132232),
     rgb(30)=(0.116656, 0.047574, 0.272321),
     rgb(45)=(0.217949, 0.036615, 0.383522),
     rgb(60)=(0.316282, 0.053490, 0.425116),
     rgb(75)=(0.410113, 0.087896, 0.433098),
     rgb(90)=(0.503493, 0.121575, 0.423356),
     rgb(105)=(0.596940, 0.154848, 0.398125),
     rgb(120)=(0.688653, 0.192239, 0.357603),
     rgb(135)=(0.775059, 0.239667, 0.303526),
     rgb(150)=(0.851384, 0.302260, 0.239636),
     rgb(165)=(0.912966, 0.381636, 0.169755),
     rgb(180)=(0.956852, 0.475356, 0.094695),
     rgb(195)=(0.981895, 0.579392, 0.026250),
     rgb(210)=(0.987464, 0.690366, 0.079990),
     rgb(225)=(0.973088, 0.805409, 0.216877),
     rgb(240)=(0.947594, 0.917399, 0.410665),
     rgb(255)=(0.988362, 0.998364, 0.644924),
     },
}

\usetikzlibrary{external}
\usepackage{subcaption}
\usepackage[hidelinks]{hyperref}
\usepackage{orcidlink}
\begin{document}

\title{Quantum time-marching algorithms for solving linear transport problems including boundary conditions}

\author{Sergio Bengoechea\orcidlink{0009-0001-8205-5878}}
\email[Corresponding author: ]{sergio.bengoechea@tuhh.de}
\affiliation{
Institute for Fluid Dynamics and Ship Theory, \href{https://ror.org/04bs1pb34}{Hamburg University of Technology}, Hamburg D-21073, Germany.
}
\author{Paul Over\orcidlink{0000-0001-7436-5254}}
\affiliation{
Institute for Fluid Dynamics and Ship Theory, \href{https://ror.org/04bs1pb34}{Hamburg University of Technology}, Hamburg D-21073, Germany.
}
\author{Thomas Rung\orcidlink{0000-0002-3454-1804}}
\affiliation{
Institute for Fluid Dynamics and Ship Theory, \href{https://ror.org/04bs1pb34}{Hamburg University of Technology}, Hamburg D-21073, Germany.
}

\date{\today}

\begin{abstract}
This article presents the first complete application of a quantum time-marching algorithm for simulating multidimensional linear transport phenomena with arbitrary boundaries, whereby the success probabilities are problem intrinsic. The method adapts the linear combination of unitaries algorithm to block encode the diffusive dynamics, while arbitrary boundary conditions are enforced by the method of images only at the cost of one additional qubit per spatial dimension. As an alternative to the non-periodic reflection, the direct encoding of Neumann conditions by the unitary decomposition of the discrete time-marching operator is proposed. 
All presented algorithms indicate optimal success probabilities while maintaining linear time complexity, thereby securing the practical applicability of the quantum algorithm on fault-tolerant quantum computers. The proposed time-marching method is demonstrated through state-vector simulations of the heat equation in combination with Neumann, Dirichlet, and mixed boundary conditions. 
\end{abstract}

\maketitle


\section{Introduction}
\label{sec:intro}
Thermofluid dynamic phenomena appear in a large range of applications in industry and academia, covering heat exchangers in power plants~\cite{Bhutta2012}, aerodynamic design in aerospace engineering~\cite{Jameson1988}, cooling systems in electronics~\cite{Zhang2021}, or atmospheric and oceanic circulation modeling~\cite{Vallis2017}. In general, such problems are solved numerically, since experimental investigations usually suffer from scaling effects and access to expensive testing facilities is often limited. However, an in-depth analysis requires extensive high-performance computing efforts, leading to greater demand for computational power as the spatio-temporal resolution increases. Alongside the challenges of further reducing the transistor size~\cite{Khan2018}, classical Central Processor Unit (CPU) technology is approaching its capacity limits to provide the scale resolution required for these applications. 
Likewise, comparable scaling constraints apply to general-purpose computing on Graphics Processing Units (GPUs)~\cite{Esmaeilzadeh2011}. 
With the objective of resolving smaller length scales~\cite{Nasa2030}, Quantum Computers (QCs) offer a promising alternative, comprising greater computational resources and more energy-efficient hardware in comparison to the classical counterparts~\cite{Jaksch2023,Givi2020}. The improvements brought about by QCs are based on an exponentially large vector space for computation and the ability to apply operations simultaneously across the entire quantum register~\cite{Nielsen2010}. 
Nonetheless, the development of QCs confronts significant challenges to establish quantum advantage, including restricted gate depth, vulnerability to noise, state preparation, or computation readout, to name a few~\cite{Preskill2018}. 

In the current era of Noisy-Intermediate-Scale-Quantum (NISQ) devices~\cite{Preskill2018}, Variational Quantum Algorithms (VQAs) have emerged~\cite{Cerezo2021} as a viable hybrid classical-quantum strategy. VQAs mitigate the I/O problem of QCs by encoding the solution in the parameters of an ansatz and updating them by measuring a single ancilla qubit~\cite{Lubasch2020}. Despite these advantageous properties, VQA methods rely on solving non-convex optimization problems, limiting the scalability of such methods and preventing --for now-- industrial applications for Computational Fluid Dynamics (CFD) \cite{Over2024a,Bengoechea2025}. 

To extend the applicability of QCs to engineering problems, the advent of fault-tolerant machines opens the door for the adaptation of classical time-marching algorithms into the quantum framework \cite{Sato2024,Alipanah2025}. 
To this end, Hamiltonian simulations, used to solve the Schr\"odinger equation \cite{Nielsen2010}, are applied to linear \cite{Brearly2024,Over2024b,An2023,Costa2019} and nonlinear Partial Differential Equations (PDEs) \cite{Esmaeilifar2024}. The advantage of this approach is to recover the PDE time evolution using the Schr\"odinger equation, as it is the governing model on a QC. For example, linear transport problems described by the wave equation can be solved under internal and external Dirichlet and Neumann boundary conditions as proposed by \citeauthor{Costa2019}~\cite{Costa2019}.

The change in methodology increases the requirements for the probabilistic design of the quantum algorithm, making high success probabilities essential to guarantee its practical feasibility. This aspect has proven to be more critical than nonlinearity implementations, as these have already been addressed either directly \cite{Lubasch2020,Esmaeilifar2024,Pool2024} or via Carleman linearization \cite{Sanavio2024}. The main challenge lies in modeling dissipation, which breaks the unitariness of quantum operations due to its irreversible nature and thereby penalizes the success probabilities. Consequently, any sequential application of non-unitary operations leads to an exponential decay in the cumulative success probability \cite{Bharadwaj2024}, making the direct implementation on a QC impractical without additional techniques \cite{Fang2023}, such as uniform singular value amplification \cite{Gilyen2019} at each time step to bound the success probabilities. However, these measures introduce an unfavorable gate scaling, making the method, at best, quadratic in the simulation time. An example of an approach with nearly certain success probabilities was proposed by \citeauthor{Brearly2024}~\cite{Brearly2024}, where the discrete time-marching operator of the advection equation was block encoded \cite{Camps2024} in the Hamiltonian simulation to permit central and upwind-biased discretization schemes, as well as boundaries. The related work of \citeauthor{Over2024b}~\cite{Over2024b} suggests an extension including a diffusive term to recover the advection-diffusion equation while maintaining optimal success probabilities with linear time complexity. 
Another methodology is presented by \citeauthor{An2023}~\cite{An2023}, where a linear combination of Hamiltonian simulations is used to approximate the non-unitary dynamics described by general Ordinary Differential Equations (ODEs). The works of \citeauthor{Jin2023}~\cite{Jin2023,Jin2024,Jin2024b,Jin2024c} and \citeauthor{Lu2024}~\cite{Lu2024} employ a different technique, called Schr\"odingerization, adapting a discrete ODE into a dilated system of Schr\"odinger equations, that is subsequently propagated in time via a Hamiltonian simulation on a QC.

The present publication introduces a novel alternative for treating non-unitary diffusive effects, maintaining optimal success probabilities, without the need for additional techniques \cite{Fang2023,Gilyen2019}. The work particularly focuses on the applicability and feasibility of the quantum time-marching algorithm proposed by \citeauthor{Over2024b}~\cite{Over2024b} to deal with engineering boundary conditions when solving time-dependent linear transport phenomena in multiple dimensions, such as heat conduction. The strategy applies the Linear Combination of Unitaries (LCU) algorithm for the operator of the governing second-order linear PDE, which is unitarily decomposed for diffusion problems. To incorporate arbitrary boundary conditions, the method of images associates the boundary type to the symmetry of the solution over the non-periodic dimension \cite{Childs2021}. Using this domain decomposition technique, the boundaries can be reconstructed from adjacent interior points. 
An alternative algorithm is also provided for diffusion problems with Neumann conditions. In this algorithm, the bounded time-marching operator is decomposed into unitaries that are directly recombined using the LCU method, eliminating the need for spatial reflection. Both proposed methods to realize boundary conditions retain optimal success probabilities while maintaining linear time complexity. This property of the algorithms is of paramount importance for their applicability in CFD simulations, as satisfying the stability criteria for explicitly solving PDEs is associated with small time steps and, therefore, a large number of them.

The paper is organized as follows. The mathematical model is described in Sec.~\ref{sec:math}. Section~\ref{sec:Q_algorithm} provides a brief review of the quantum methods required to implement discrete operations, such as those presented in Sec.~\ref{sec:math} on a fault-tolerant QC, along with the corresponding quantum circuits. Two-dimensional state-vector simulations of the heat equation with all types of boundary conditions are benchmarked against classical Finite Differences (FD) solutions in Sec.~\ref{sec:simulations}. 
Conclusions and future directions are summarized in Sec.~\ref{sec:conclusions}. Additionally, the Appendix~\ref{sec:appendix} illustrates a matrix-circuit analogy to facilitate the understanding of the quantum circuits, with particular emphasis on the LCU algorithm. 

Throughout the publication, continuous vectors are indicated as $\vec{\square}$, discrete vectors in \textbf{bold}, quantum states in the Dirac (or bra-ket) notation, while matrices and gates are marked with capital letters. Furthermore, the \textit{big-endian} convention applies to all quantum circuits and registers.

\section{Mathematical Model}
\label{sec:math}
Consider a two-dimensional ($d=2$) domain, with spatial coordinates $\vec{x} \in [0,1]\times[0,1]$~{$\text{[m]}$}, the time $t \in [0,T)$~{$\text{[s]}$}, a generic transport scalar $\phi$~{$\text{[-]}$}, a (constant) dynamic viscosity ${\Gamma\;[\text{m}^2\text{/s}]}$ and the dynamics given by the unsteady diffusion equation, e.g., the heat conduction, viz. 
\begin{equation}
    \frac{\partial \phi}{\partial t} - \Gamma \, \nabla^2 \phi = 0 \, . 
    \label{eq:heat_equation}
\end{equation}

This parabolic equation is discretized using an isotropic and equidistant spatial $\Delta x$~{$\text{[m]}$} and temporal $\Delta t$~{$\text{[s]}$} spacing with $N = N_{\text{x}_1}^d$ and $N_{\text{t}}$ supporting points, respectively. 
For the sake of clarity, the discretization of Eqn.~\eqref{eq:heat_equation} is demonstrated for a one-dimensional ($d=1$) problem with $N_{\text{x}_1}=\sqrt[d]{N}$ supporting points. The differential operators are approximated with a first-order accurate explicit forward Euler method in time and second-order central differences for the spatial derivative. Moreover, periodic boundary conditions are assumed for simplicity. Accordingly, the discrete form of Eqn.~\eqref{eq:heat_equation} is
\begin{equation}
\frac{\phi^{t+1}_j-\phi^t_j}{\Delta t} - \Gamma\frac{\phi^{t}_{j+1} - 2\phi^t_j + \phi^{t}_{j-1}}{\Delta x^2} = 0 \, .
\end{equation}
Here, the Toeplitz matrix $A \in \mathbb{R}^{N_{\text{x}_1} \times N_{\text{x}_1}}$, viz. 
\begin{equation}
    A = 
    \begin{bmatrix}
        1{-} \, 2r_\text{h}  & r_\text{h}      &        & 0        & r_\text{h}  \\
        r_\text{h}      & 1{-} \, 2r_\text{h} & \ddots &          &  0   \\
                 & \ddots   & \ddots & \ddots   &      \\
        0        &          & \ddots & 1{-} \, 2r_\text{h} & r_\text{h}  \\
        r_\text{h}  & 0        &        & r_\text{h}      & 1{-} \, 2r_\text{h}
    \end{bmatrix} \, ,
    \label{eq:A_matrix}
\end{equation}
is the time-marching operator in $\pmb{\phi}(t+\Delta t) = A \pmb{\phi}(t)$, where the non-dimensional parameter $r_\text{h}=\Gamma\,\Delta t/\Delta x^2$~{$\text{[-]}$} characterizes the spatio-temporal behavior of the diffusive transport. 
Following the time-marching idea, Eqn.~\eqref{eq:heat_equation} is solved sequentially in time, with the time step only being restricted by the stability criteria determined from the magnitude of $r_\text{h}$. This defines a convergence condition of the numerical method applied to the ODE, requiring values below one in all supporting points for the duration of the entire solution process.

\begin{figure}[htpb]
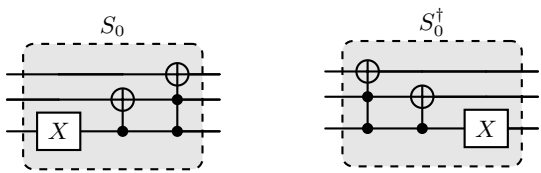

    \centering
    \begin{subfigure}[b]{0.23\textwidth}
        \centering
        \include{Figures/01-LCU-Diff_S0}
        \caption{Quantum circuit for $S_0$.}
        \label{fig:lcu:diff_S0}
    \end{subfigure}
    \begin{subfigure}[b]{0.23\textwidth}
        \centering
        \include{Figures/01-LCU-Diff_S0Dagger}
        \caption{Quantum circuit for $S_0^\dagger$.}
        \label{fig:lcu:diff_S0dagger}
    \end{subfigure}
    \caption{Three-qubit example for the implementation of the shift operators $S_0$ and $S_0^\dagger$.}
    \label{fig:lcu:diff_S0S0dagger}
\end{figure}

Prior to its application on a QC, the matrix $A$ is subjected to modifications.
First, the main diagonal is rescaled to one by $A/(1-2 \, r_\text{h})$. 
Second, this scaled matrix is decomposed w.r.t. its non-zero diagonals as 
\begin{equation}
    \frac{A}{1-2 \,r_\text{h}} = I + \frac{r_\text{h}}{1-2 \,r_\text{h}}(S_0+S_0^{\dagger}) \, , 
    \label{eq:A_decomp}
\end{equation}
using the unitaries identity $I$ and the shift operator
\begin{equation}
    S_0 =
    \begin{bmatrix}
        0      & 1      & 0      & \cdots &  0  \\
               & 0 & 1 &        &  \vdots   \\
        \vdots &        & \ddots & \ddots &  0   \\
        0      &        &        & 0 &  1   \\
        1      & 0      & \cdots &        &  0
    \end{bmatrix}\,.
    \label{eq:S0_operator}
\end{equation}
In Eqn.~\eqref{eq:A_decomp}, $S_0^{\dagger}$ indicates the conjugate transpose of $S_0$, and both of them can be efficiently implemented with the sequence of quantum gates depicted in Fig.~\ref{fig:lcu:diff_S0S0dagger}. 

\section{Quantum Methods and Algorithms}
\label{sec:Q_algorithm}
This section introduces the fundamental methodologies used to implement the discrete operators of Sec.~\ref{sec:math} on the QC. 
These are amplitude encoding (Sec.~\ref{sec:Amp_enco}), Hamiltonian simulation (Sec.~\ref{sec:Ham_sim}), block encoding (Sec.~\ref{sec:Block_encod}), and the LCU algorithm (Sec.~\ref{sec:lcu}). 
Subsequently, the algorithmic framework is described, with the focal point being the treatment of boundary conditions, cf. Sec.~\ref{sec:boundaries}. 

\subsection{Amplitude encoding}
\label{sec:Amp_enco}
Unlike classical computation, where information is represented by binary strings in the computational basis, arbitrary variable functions can be encoded in the amplitudes $\psi_j \in \mathbb{C}$ of a quantum state $\ket{\psi} \in \mathbb{C}^N$. 
Here, the Dirac (or bra-ket) notation, standard in quantum mechanics, is used. In this notation, a normalized column vector $\pmb{\psi}$ is denoted as $\ket{\psi}$ and its conjugate transpose reads $\ket{\psi}^\dagger =(\ket{\psi}^\intercal)^*  = \bra{\psi}$ \cite{Griffiths2018}.
Note that the bra-ket notation ($\bra{\cdot}$ or $\ket{\cdot}$) always indicates a normalized quantum state. For a quantum register consisting of $n=\log_2(N)$ qubits, the state vector of size $N$ is thus given as 
\begin{equation}
    \ket{\psi} = \sum_{j=0}^{N-1} \psi_j \ket{e_j} \, ,
    \label{eq:Amp_enco}
\end{equation}
where $\ket{e_j}$ form an orthonormal basis of a Hilbert space and $\psi_j^2$ represents the probability of measuring the $j$-th basis state $\ket{e_j}$ conserving $\sum_j \psi_j^2=1$. 
Operations such as the (inner) scalar and (outer) dyadic product can be compactly expressed using the bra-ket notation as $\langle \cdot |\cdot \rangle$ and $\ket{\cdot}\bra{\cdot}$, respectively. In the special case of concatenating qubit states represented as two-column vectors or two-row vectors, the Kronecker product $\otimes$ exemplarily given for an $n=2$ qubit system is
\begin{equation}
\begin{split}
 \bra{\rho \psi} &= \bra{\rho} \otimes \bra{\psi}  \\
 &=
\begin{bmatrix}
\rho_1 \psi_1,  & \rho_1 \psi_2, & 
\rho_2 \psi_1,  & \rho_2 \psi_2   
\end{bmatrix} \\
&= (\ket{\rho} \otimes \ket{\psi} )^\dagger = \ket{\rho\psi}^\dagger \,.
\end{split}
\end{equation}

In general, the state vector is prepared either by initializing the amplitudes $\psi_j$ through a learning approach \cite{Over2024a,Bengoechea2025}, by employing divide-and-conquer strategies \cite{Esmaeilifar2024}, by programming through Tensor Trains (TTs) \cite{Ran2020,Zhou2021,Termanova2024}, or in special cases by using analytical approaches \cite{Oseledets2013}. 

\subsection{Hamiltonian simulation}
\label{sec:Ham_sim}
\begin{figure}[htpb]
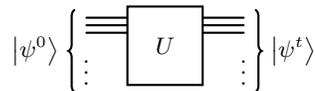

    \centering
    \include{Figures/02-Ham_Sim}
    \caption{Generic circuit for the temporal evolution of $\ket{\psi^0}$ given by the Hamiltonian simulation of Eqn.~\eqref{eq:schroedinger}.}
    \label{fig:ham_sim}
\end{figure}

The Hamiltonian dynamics of a wave function $\psi(\vec{x},t)$ are described by the time-dependent Schr\"odinger equation, 
\begin{equation}
    {i}\frac{d \psi(\vec{x},t)}{dt} = \hat H \psi(\vec{x},t) \, , \quad \psi(\vec{x},0)=\psi^0 \, .
    \label{eq:schroedinger}
\end{equation}
This ODE represents a linear and dissipation-free (reversible) system, with the Hamiltonian $\hat H$ being Hermitian, $\hat H=\hat H^{\dagger}$, and ${i}=\sqrt{-1}$ denoting the imaginary unit. For a time-independent $\hat H$, the solution to Eqn.~\eqref{eq:schroedinger} is given by $\psi(\vec{x},t) = \psi^0(\vec{x}) \, \exp(-{i}\hat H t)$ \cite{Nielsen2010,Griffiths2018}. Given the fact that the operator $\hat U=\exp(-{i}\hat Ht)$ is unitary, a direct implementation of an equivalent quantum circuit exists, providing a deterministic implementation on a QC. The discretized Hamiltonian operator $H \in \mathbb{C}^{N \times N}$ is applied to the spatially discretized initial wave function $\pmb{\psi}^0 \in \mathbb{C}^{N}$, encoded by the initial quantum state $\ket{\psi^0}$, to approximate its temporal evolution by $\ket{\psi^t} = U \ket{\psi^0}$ \cite{Berry2015}, as it is depicted in Fig.~\ref{fig:ham_sim}. 

There exist multiple algorithms based on this quantum state evolution \cite{Childs2018}, such as product formulas to expand a unitary evolution (e.g., Suzuki-Trotter \cite{Hatano2005}), Taylor series \cite{Berry2015_Taylor}, quantum signal processing \cite{Low2017}, or the Schr\"odingerization process \cite{Jin2023,Jin2024,Jin2024b} in which the underlying PDE is cast into a dilated system of Schr\"odinger equations~\eqref{eq:schroedinger} to be solved using Hamiltonian simulations. A specific application of the Hamiltonian simulation is described in \cite{Brearly2024, Costa2019, Gingrich2004} where this technique is employed to block encode non-unitary operators, cf. Sec.~\ref{sec:Block_encod}. 

\subsection{Block encoding}
\label{sec:Block_encod}
Applying FD approximations to PDEs results in discrete operators $M$, e.g., matrix $A$ in Sec.~\ref{sec:math}. These matrices are generally non-unitary, so that $M^\dagger M \neq I$ and $M M^\dagger \neq I$, which poses a challenge when implementing them on a QC. Since quantum dynamics must be expressed through unitary operations to preserve norm and probability amplitudes, a common approach is to enlarge the underlying vector space and the operator matrix for a unitary embedding, cf. \cite{Camps2024}, according to 
\begin{equation}
   U_{\text{M}} =  \begin{bmatrix}
        M/\alpha & \ast\\
        \ast & \ast 
    \end{bmatrix} \,,
    \label{eq:block_encoding}
\end{equation}
where the sub-blocks indicated with $\ast$ are selected differently. Note that the unitary matrix $U_{\text{M}}$ exists only if the singular values ($\sigma_r$) of any submatrix block are bounded by $1$, i.e., its spectral norm $\|M\|_2 = \sigma_{\text{max}}(M)\leq1$ \cite{Camps2024}, where $\sigma_{\text{max}}$ is the largest singular value. 
This implies that the l2-norm of each column of $M$ must not exceed $1$.
To satisfy this condition, $M$ is pre-scaled by a normalization factor $\alpha \, \in \mathbb{R^+}$ such that $\alpha \geq \|M\|_2$. 
Consequently, for the non-unitary matrix $M \in \mathbb{C}^{2^n\times2^n}$ defined by $n$ qubits, an accurate block encoding is achieved primarily by an optimal choice of the subnormalization factor, i.e., $\alpha = \|M\|_2$. For the sake of clarity, $\tilde{M}$ is assumed to be the optimal scaling of $M$, i.e., $\tilde M = M/\alpha$. In general, the scaling increases the condition number of $\tilde M$ \cite{LapworthSuenderhauf2025}, making block-encoding strategies for quantum linear system solvers not practical. In contrast, time-marching approaches do not encounter this problem due to the absence of matrix inversion. 

To complete the block encoding, it remains to select the $\ast$ entries in Eqn.~\eqref{eq:block_encoding}. This choice is not unique. A general method proposed by \citeauthor{Camps2024}~\cite{Camps2024} uses non-parameterized sub-blocks for $U_{\text{M}}$ as 
\begin{equation}
   U_{\text{M}} =\begin{bmatrix} 
   \tilde{M}                              & \sqrt{I- \tilde M^\dagger \tilde{M}} \\
   \sqrt{I - \tilde{M}^\dagger \tilde{M}} &  -\tilde{M}
    \end{bmatrix}  \, .
    \label{eq:block_encoding_Camps}
\end{equation}
An alternative to compute the square root of $\tilde{M}^\dagger \tilde{M}$ in Eqn.~\eqref{eq:block_encoding_Camps} is also provided in Ref.~\cite{Camps2024}, where the appearance of this term is avoided in the sub-blocks of $U_\text{M}$. 
This approach is tailor-made for certain dynamics that recover Toeplitz matrices with constant non-zero diagonals upon discretization, and it requires more than one ancilla qubit. 

A numerically motivated block encoding has been introduced by \citeauthor{Lin2021}~\cite{Lin2021}, which determines the sub-blocks of the structured matrix
\begin{equation}
U_{\text{M}} =\begin{bmatrix}
            sM & G \\
            F & W \,
\end{bmatrix} \, ,
\label{eq:block_encoding_Lin}
\end{equation}
imposing $F^\dagger F = I - s^2 M^\dagger M$ and $FF^\dagger = I - s^2 M M^\dagger$ for $U_{\text{M}}$ to be unitary with $s=1/\sigma_{\text{max}}(M)$. 
The remaining sub-blocks, $G$ and $W$, are numerically constructed using $QR$ decomposition on randomly generated matrices of the same size. Meanwhile, the first $2^N$ columns of $U_{\text{M}}$ remain unaffected by the $QR$-based operation since these are orthogonal. 

\begin{figure}[htpb]
    \centering
    \input{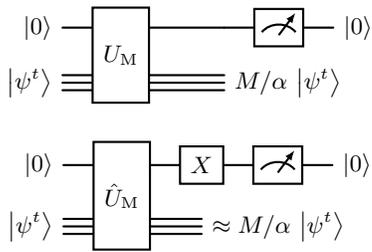}
    \caption{Exemplary quantum circuit implementations of the block encoding strategies for $U_{\text{M}}$ (top) and $\hat{U}_{\text{M}}$ (bottom).}
    \label{fig:Ums}
\end{figure}

A different block encoding for time-marching operators derived from pure advective transport is based on the Hamiltonian simulation, cf. Sec.~\ref{sec:Ham_sim}. This procedure follows Ref.~\cite{Brearly2024} which encodes the non-unitary matrix $\tilde{M}$ as 
\begin{equation}
\begin{split}
    \hat{U}_{\text{M}} = \begin{bmatrix}
        \ast      & \ast\\
        \tilde{M} & \ast 
    \end{bmatrix}= \exp(-i H \theta)&, \\
    \text{with}& \; 
    H = \begin{bmatrix}
        0              & -i \, \tilde{M}^\dagger \\
        i \, \tilde{M} &  0 
    \end{bmatrix}\, .
\end{split}
    \label{eq:Ham_sim_BE}
\end{equation}
\citeauthor{Brearly2024}~\cite{Brearly2024} propose to transform Eq.~\eqref{eq:Ham_sim_BE} by applying trigonometric identities \cite{Gingrich2004} as follows 
\begin{equation}
   \hat{U}_{\text{M}} =\begin{bmatrix} 
        \cos(\sqrt{\tilde{M}^\dagger  \tilde{M}}\, \theta)  &  - \tilde{M}^\dagger\frac{\sin(\sqrt{\tilde{M} \tilde{M}^\dagger}\, \theta)}{\sqrt{\tilde{M}  \tilde{M}^\dagger}} \\
        \tilde{M} \frac{\sin(\sqrt{\tilde{M}^\dagger \tilde{M}} \, \theta)}{\sqrt{\tilde{M}^\dagger \tilde{M}}}  &  \cos(\sqrt{ \tilde{M} \tilde{M}^\dagger}\,\theta) 
    \end{bmatrix} \, .
    \label{eq:brearley_encoding}
\end{equation}
In Eqns.~\eqref{eq:Ham_sim_BE} and \eqref{eq:brearley_encoding}, the control parameter $\theta \in \mathbb{R}$ denotes a Hamiltonian evolution time, defining both the accuracy of the block encoding and its probability of success. Equation~\eqref{eq:brearley_encoding} is used to temporarily evolve the advection dynamics by applying the unitary operator $\hat{U}_{\text{M}}$ in \cite{Brearly2024}. The application of $\hat{U}_{\text{M}}$ to the prepared quantum state $\ket{\psi^t}$, indicated at the bottom of Fig.~\ref{fig:Ums}, implements
\begin{equation}
    \hat{U}_{\text{M}} \ket{0} \ket{\psi^t} = \begin{bmatrix}
        \tilde{M} \ket{\psi^t} \\
        I \ket{\psi^t}
    \end{bmatrix} \approx  \begin{bmatrix}
        \ket{\psi^{t+1}}\\
        \ket{\psi^t} 
    \end{bmatrix} \, .
    \label{eq:q_adv_state}
\end{equation}
In the post-selection step, either the updated state $\ket{\psi^{t+1}}$ or the corresponding  initial state $\ket{\psi^t}$ is measured in the computational basis $\ket{0}$ or $\ket{1}$, respectively. This choice is governed by the success probability per time step $p^t$, which for $\theta=0$ results in zero and for $\theta={\pi}/{2}$, the optimal value of $p^t \approx 1$ is reached. 

Mind that $U_{\text{M}}$ and $\hat{U}_{\text{M}}$ differ in structure with the block encoding of $\tilde{M}$ located in the upper-left and lower-left sub-blocks, respectively. This distinction must be considered for the circuits' implementation and the post-selection output as it is depicted in Fig.~\ref{fig:Ums}. 

\smallskip
For all block encoding strategies, the choice of $\ast$ determines the success probabilities associated with it, as the method executes a non-deterministic operation on the QC, and introduces a penalty to the overall success probability. 
For example, the application of $U_{\text{M}}$ to a quantum register initialized in the state $\ket{0} \otimes \ket{\psi(\vec{x},t)}$ evolves the system to 
\begin{equation}
    \ket{0}\big(M/\alpha\ket{\psi(\vec{x},t)} + \ast \, \mathbf{0} \big) + \ket{1}\big( \ast \ket{0}\ket{\psi(\vec{x},t)} \big) \, . 
\end{equation}
Here, the term $\ast \, \mathbf{0}$ indicates the action of the upper-right sub-block in Eqn.~\eqref{eq:block_encoding} on the lower half of the register, while $\ket{1}\,\big( \ast \ket{0}\ket{\psi(\vec{x},t)} \big)$ corresponds to the lower row in Eqn.~\eqref{eq:block_encoding}. 
This transformation is accurately implemented up to the subnormalization factor $\alpha = \|M\|_2$, conditioned on measuring $\ket{0}$ on the additional (ancilla) qubit. 
Thus, the block encoding prepares the target state $M/\alpha \ket{\psi}$ with a success probability of 
\begin{equation}
\begin{split}
     p &= (1/\alpha^2) \bra{\psi}  M^\dagger M \ket{\psi} 
     = 1/\alpha^2 \|M \ket{\psi}\|_2^2 \\
     &\leq 1/\alpha^2 \|M\|_2^2 \|\ket{\psi}\|_2^2 
     = 1/\alpha^2 \|M\|_2^2\,  .
\end{split}
\label{eq:succ_prob}
\end{equation}
For $\alpha>1$, the cumulative success probability of any time-marching approach successively decreases with each temporal evolution step. The success probabilities rapidly converge to zero, making the algorithm impractical except for $\alpha = 1$ \cite{Over2024b}. In general, any matrix can be block encoded. However, for nearly unitary matrices when $\alpha \approx 1$, the success probability approaches one. This is the case for the block encoding strategy suggested by \citeauthor{Brearly2024}~\cite{Brearly2024}, where the parameter $\theta$ can be selected to obtain a success probability near unity. In probabilistic terms, the strategy proposed in Ref.~\cite{Brearly2024}, cf. Eqns.~\eqref{eq:Ham_sim_BE} and (\ref{eq:brearley_encoding}), has clear advantages over the block encoding strategies in Eqns.~\eqref{eq:block_encoding_Camps} and~\eqref{eq:block_encoding_Lin}. 

\citeauthor{Fang2023}~\cite{Fang2023} proposed using Quantum Singular Value Transformation (QSVT) to correct vanishing probabilities, thereby restoring practical feasibility. However, the proposed modification comes with an increase from linear to quadratic complexity in time. 
In the next sections, further alternatives to QSVT, maintaining $\alpha = 1$ are exploited while preserving the linear time complexity. 

\subsection{Linear combination of unitaries}
\label{sec:lcu}
\begin{figure}[htpb]
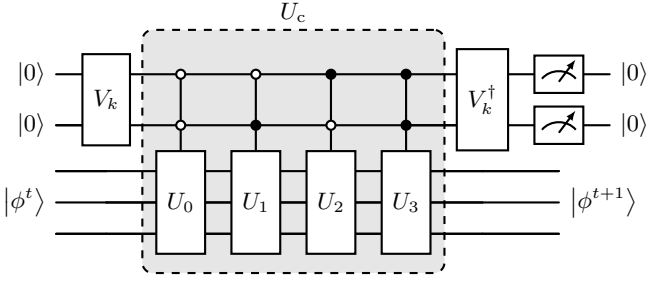

        \centering
        \include{Figures/04-LCU_Heat_1D}
        \caption{Quantum circuit to implement one explicit time step of the one-dimensional periodic heat equation, Eqn.~\eqref{eq:heat_equation}.}
        \label{fig:lcu:onedim}
\end{figure}

\begin{figure*}[htbp]
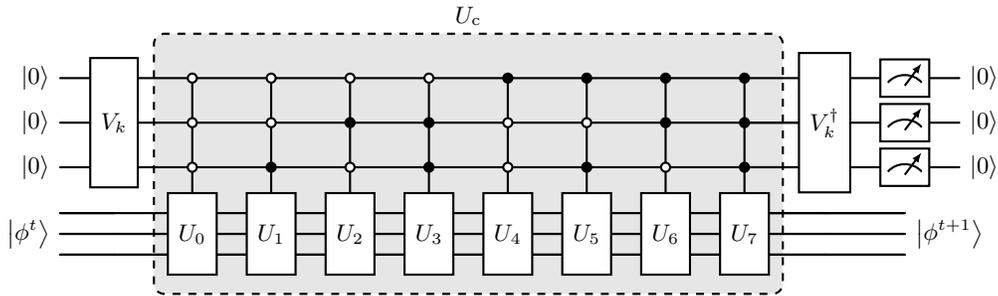

    \centering
    \include{Figures/05-LCU_2D}
    \caption{LCU circuit for the two-dimensional periodic heat equation.}
    \label{fig:lcu:twodim}
\end{figure*}

The Linear Combination of Unitaries (LCU) algorithm combines a weighted sum of $m$ unitary operators $U_k$ on a QC \cite{Childs2012}. These unitary operators are not necessarily obtained through the block encoding strategies applied to $U_{\text{M}}$ in Sec.~\ref{sec:Block_encod}. With this approach, a decomposition into unitaries of a non-unitary matrix 
\begin{equation}
    M = \sum_{k=0}^{m-1} \kappa_k U_k \, ,
    \label{eq:M_decomp_lcu}
\end{equation}
with $\kappa_k \in \mathbb{R}^+ \cup \{0\}$, can be implemented with the circuits of Fig.~\ref{fig:lcu:onedim}, or Fig.~\ref{fig:lcu:twodim} for the example of ${m=4}$, and ${m=8}$ unitaries, respectively. These quantum circuits prepare the final state $\ket{\phi^{t+1}}$, using $n_{\text{anc.}} = log_2(m)$ ancilla qubits, conditioned to measure $\ket{0}^{\otimes n_{\text{anc.}}}$, viz. 
\begin{equation}
    \bra{0} \underbrace{V^{\dagger} U_\text{c} V}_{U_{\text{M}}} \ket{0}\ket{\phi^t}   = \frac{M}{\alpha} \ket{\phi^t} \, .
    \label{eq:LCU_BE}
\end{equation} 
Comparing Eqn.~\eqref{eq:LCU_BE} with Eqn.~\eqref{eq:block_encoding} from  Sec.~\ref{sec:Block_encod}, the LCU can be interpreted as a block encoding strategy where $\alpha$ is the scaling factor and $1/\alpha^2$ is the penalty in the success probability, as indicated in  Eqn.~\eqref{eq:succ_prob}. 
The LCU method can also be seen as the implementation of a summation on a QC. Since this operation is not norm-preserving and, in general, cannot be performed deterministically, it is reflected in losses in the success probability. 

In the application of the LCU algorithm, the first step is to decompose the arbitrary matrix into unitaries. To this end, the Pauli decomposition employs the complete basis formed by Pauli matrices (or their tensor products) to represent the operators acting on $n$ qubits \cite{Hantzko2024}. Once the unitary decomposition of $M$ is achieved, each resulting unitary operator $U_k$ is implemented in the circuits of Fig.~\ref{fig:lcu:onedim} or Fig.~\ref{fig:lcu:twodim}, and it can be understood as a block matrix on the main diagonal of a larger matrix space. Their ordering on the diagonal matrix is defined by the binary selection of the controlled ancilla qubits, being $\ket{00}$ the first diagonal block matrix element, $\ket{01}$ the second diagonal block, etc.
Appendix~\ref{sec:appendix} describes the mapping between quantum gates and matrices for the quantum circuit of Fig.~\ref{fig:lcu:onedim} in greater detail. The weights $\kappa_k$ are contained in the first column of $V_k$ as modified coefficients. Any excess unitaries in the power of two can easily be discarded by setting the corresponding $\kappa_k$ coefficient in $V_k$ to zero.  Exemplified for ${m=4}$, the unitary $V_k$ reads 
\begin{equation}
    V_k = 
    \begin{bmatrix}
        \sqrt{\frac{\kappa_0}{\kappa_0+\kappa_1+\kappa_2+\kappa_3}} &  \\
        \sqrt{\frac{\kappa_1}{\kappa_0+\kappa_1+\kappa_2+\kappa_3}} & \dots \\
        \sqrt{\frac{\kappa_2}{\kappa_0+\kappa_1+\kappa_2+\kappa_3}} & \dots \\
        \sqrt{\frac{\kappa_3}{\kappa_0+\kappa_1+\kappa_2+\kappa_3}} & 
    \end{bmatrix}\,,
\end{equation}
with the square roots appearing due to normalization, and the rest of the columns ($\dots$) forming an orthonormal basis to make $V_k$ unitary. This could be achieved by the application of the Gram-Schmidt process on randomly initialized columns.  

In particular, the circuit of Fig.~\ref{fig:lcu:onedim} prepares the initial state as 
\begin{equation}
    \begin{split}
    \ket{00}\ket{\phi^t}  \to& 
    \ket{00}\tilde{A}_{00}\ket{\phi^t} + \ket{01} \tilde{A}_{01} \ket{\phi^t} \\
    & \quad + \ket{10} \tilde{A}_{10} \ket{\phi^t} + \ket{11} \tilde{A}_{11} \ket{\phi^t} \, ,
    \label{eq:mapping_1d}
    \end{split}
\end{equation}
with the matrices
\begin{align}
    \label{eq:A_00_1d}
    \tilde{A}_{00} &=\frac{ \kappa_0 U_0 + \kappa_1 U_1 + \kappa_2 U_2 + \kappa_3 U_3 }{\kappa_0 + \kappa_1 + \kappa_2 + \kappa_3}\\
    \label{eq:B_01_1d}
    \begin{split}
    \tilde{A}_{01} &= \frac{1}{\kappa_0 + \kappa_1 + \kappa_2 + \kappa_3}\bigg( \sqrt{\kappa_1 \big(\kappa_0+\kappa_2+\kappa_3 \big)} \; U_1 - \\ 
    &\sqrt{{\kappa_1}\big({\kappa_0+\kappa_2+\kappa_3}\big)^{-1}}\big(\kappa_2 U_2+\kappa_3 U_3+\kappa_0 U_0 \big) \bigg)
    \end{split}\\
    \label{eq:C_10_1d}
    \begin{split}
    \tilde{A}_{10} &= \sqrt{\frac{\kappa_2}{\big(\kappa_0+\kappa_2+\kappa_3 \big) \big( \kappa_0+\kappa_1+\kappa_2+\kappa_3 \big)}} \\ 
    &\bigg( \sqrt{\big(\kappa_0+\kappa_3 \big)} \; U_2 - \sqrt{{\big(\kappa_0+\kappa_3 \big)^{-1}}}\big( \kappa_3 U_3 + \kappa_0 U_0 \big) \bigg) 
    \end{split}\\
    \tilde{A}_{11} &= \sqrt{\frac{\kappa_0 \kappa_3}{\big(\kappa_0+\kappa_3\big) \big(\kappa_0 + \kappa_1 + \kappa_2 + \kappa_3 \big)}} \big(U_3-U_0 \big)\,.
    \label{eq:D_11_1d}
\end{align}

From Eqn.~\eqref{eq:mapping_1d}, it is evident that the state $\ket{00}\tilde{A}_{00}\ket{\phi^t}$ needs to be post-selected out of the superposition of the four computational basis states to recover the action of $\tilde{A}_{00} =M/\alpha$ with $\alpha=\sum_k \kappa_k$. 
The probability of measuring that particular basis state $p_{\ket{00}}$ is expressed by the squared Euclidean norm of the quantum state in the Hilbert space, viz.
\begin{equation}
    \begin{split}
        p_{\ket{00}} &= \bra{\phi} \tilde{A}^\dagger_{00} \tilde{A}_{00} \ket{\phi} = \|\tilde{A}_{00} \ket{\phi}\|^2_2 \leq \| \tilde{A}_{00} \|^2_2 \, \|\ket{\phi}\|^2_2 \\ =\| &\tilde{A}_{00}  \|^2_2 = \sigma_{\text{max}}(\tilde{A}_{00})^2 = \lambda_{\text{max}}(\tilde{A}^\dagger_{00} \tilde{A}_{00} )\, ,
    \end{split}
\end{equation}
where $\sigma_{\max}$ and $\lambda_{\max}$ denote the largest singular value and eigenvalue, respectively. The same applies to the measurement probabilities of the other computational basis $p_{\ket{01}}$ ($\leq \sigma_\text{max}(\tilde A_{01})^2$), $p_{\ket{10}}$ ($\leq \sigma_\text{max} (\tilde A_{10})^2$), and $p_{\ket{11}}$ ($\leq \sigma_\text{max} (\tilde A_{11})^2$). 

In principle, any unitary decomposition of $M$ is possible. However, it is crucial to ensure an optimal probability of success. To this end, the unitary decomposition given in Eqn.~\eqref{eq:A_decomp} for the time-marching solution of the one-dimensional heat equation is considered. One step of the first-order temporal evolution ${\pmb{\phi}(t+\Delta t) = A \pmb{\phi}(t)}$ can be implemented exclusively with three unitaries ($I,S_0,S_0^\dagger$) in the LCU circuit of Fig.~\ref{fig:lcu:onedim} with ${\kappa_0 = 1}$, ${\kappa_1 = \kappa_2 = r_\text{h}/(1-2r_\text{h})}$, and ${\kappa_3 = 0}$. 
Note that the additional constraint $r_\text{h} \leq 1/(2\,d)$ needs to be imposed to guarantee that all the LCU weights are $\kappa_k \geq 0$. 
Here, the LCU scaling factor ${1/(\kappa_0+\kappa_1+\kappa_2+\kappa_3) = 1-2r_\text{h}}$ cancels the initial subnormalization factor $1/(1-2r_\text{h})$ in Eqn.~\eqref{eq:A_decomp} completely and results in the exact recovering of $A$ with $\alpha=1$. Accordingly, the success probability of measuring the computational basis $\ket{00}$ for one time step is given by ${p^t= \bra{\phi^t} A^\dagger A \ket{\phi^t} = \| \pmb{\phi}^{t+1}\|_2^2/\|\pmb{\phi}^t \|_2^2}$ since the recovered operator $A$ is not unitary \cite{Over2024b}. For simulating one step of the two-dimensional heat equation, the quantum circuit in Fig.~\ref{fig:lcu:twodim} can be applied in the same manner. 

As a result, the time-marching operator $A$ can be implemented in the LCU method to resolve the temporal evolution of the parabolic PDE in a time-marching fashion without vanishing cumulative probabilities $p_{\text{c}}$ due to ${\alpha = 1}$. In particular, $p_{\text{c}}=\prod\limits^{T}_0 p^t=\| \pmb{\phi}^T\|_2^2/\|\pmb{\phi}^0 \|_2^2$ and converges for steady-state solutions, depending solely on the temporal distance between initial and final state. Therefore, it is an intrinsic property of the problem under investigation rather than of the applied numerical method.

\subsection{Non-periodic boundary conditions}
\label{sec:boundaries}
The method of images enforces boundary conditions by reflecting the bounded field and thereby exploiting the symmetric and antisymmetric properties of Neumann and Dirichlet boundaries, respectively \cite{Childs2021}. This implies a two-fold increase in the degrees of freedom per non-periodic dimension, making the approach less appealing for classical computation. However, symmetry reflections on a QC can be implemented efficiently by introducing a single qubit for each non-periodic dimension in $\mathbb{R}^d$ \cite{Sano2024}, exploiting the exponential scaling in state space of the amplitude encoding strategy \cite{Nielsen2010}. 
This permits boundary treatment through the even (sign-preserving) or odd (sign-changing) symmetry of neighboring points using discrete periodic operators, while conserving the advantage of non-vanishing success probabilities of the time-marching method, as described in Sec.~\ref{sec:lcu}. 

\begin{figure*}[htbp]
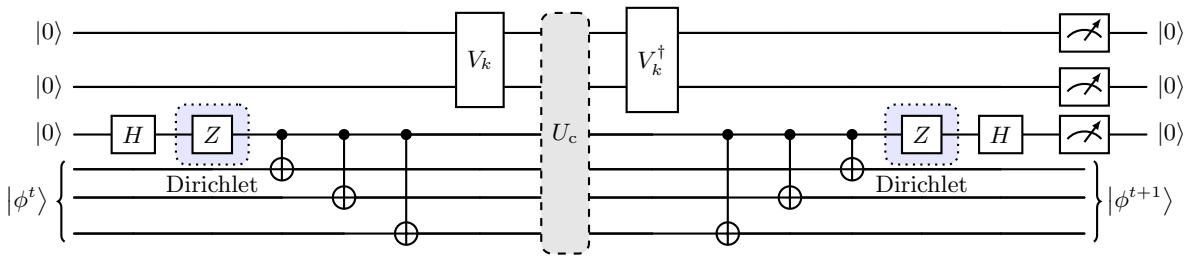

    \centering
    \include{Figures/07-PDE_Reflec_1D}
    \caption{Quantum circuit to implement one explicit time step of the one-dimensional non-periodic PDE defined by the $U_\text{c}$ and $V_k$ blocks. The total number of qubits is six with three qubits dedicated to $N=8$ supporting points. For the Neumann boundaries, the even reflection of the register is prepared exclusively with Hadamard and CNOT gates, while the odd reflection of the Dirichlet condition also includes $Z$ gates.}
    \label{fig:lcu:onedim_PDE_BC}
\end{figure*}

\begin{figure*}[htpb]
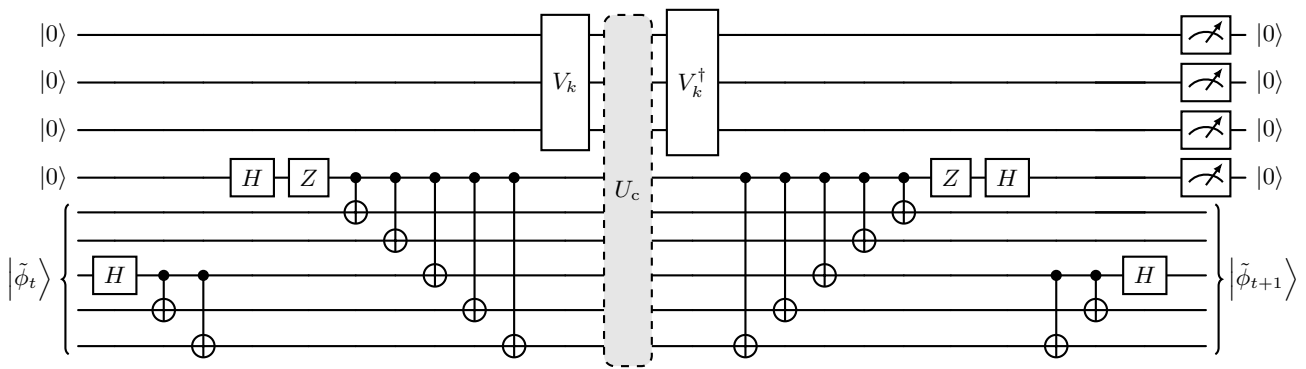

    \centering
    \include{Figures/09-PDE_Reflec_2D}
    \caption{Quantum circuit to implement one explicit time step of the two-dimensional non-periodic PDE defined by the $U_\text{c}$ and $V_k$ blocks. The total number of qubits is nine with four qubits dedicated to $N=16$ supporting points. Mixed boundaries are implemented with Neumann in the first index and Dirichlet in the second.}
    \label{fig:lcu:twodim_PDE_MixedBC}
\end{figure*}

The circuits depicted in Fig.~\ref{fig:lcu:onedim_PDE_BC}, \ref{fig:lcu:twodim_PDE_BC}, and \ref{fig:lcu:twodim_PDE_MixedBC} illustrate the implementation of non-periodic boundary conditions for a single time step. In these circuits, the initial state $\ket{0 \, \phi^t}$ is mirrored from the upper half of the register onto the lower half according to the desired boundary conditions. Specifically, these circuits implement this reflection for one-dimensional (Fig.~\ref{fig:lcu:onedim_PDE_BC}), two-dimensional (Fig.~\ref{fig:lcu:twodim_PDE_BC}), and mixed two-dimensional boundaries (Fig.~\ref{fig:lcu:twodim_PDE_MixedBC}), respectively. After executing the unitary transformations $V_k$, $U_\text{c}$, and $V_k^\dagger$ which implement one explicit time step of the quantum time-marching algorithm, the conjugate transpose of the reflection circuit is applied to uncompute the initial reflection of the prepared state, returning the quantum register to $\ket{0 \, \phi^{t+1}}$ conditioned to measure on the $\ket{0}^{\otimes n_{\text{anc.}}}$ basis. In case of two-dimensional domains, the reflection circuits require the modifications shown in Fig.~\ref{fig:lcu:twodim_PDE_BC} and \ref{fig:lcu:twodim_PDE_MixedBC} for a lexicographic order of the two-dimensional field into a one-dimensional vector $\tilde{\pmb{\phi}}{}^{t}$ or $\tilde{\pmb{\phi}}{}^{t+1}$. Note that the even (sign-preserving) symmetry found for Neumann boundaries is implemented with Hadamard and controlled-NOT (CNOT) gates \cite{Sano2024}, while the odd (sign-changing) reflection in Dirichlet boundaries employs additional $Z$ gates. 
\begin{figure*}[htpb]
    \centering
    \include{Figures/08-PDE_Reflec_2D}
    \caption{Quantum circuit to implement one explicit time step of the two-dimensional non-periodic PDE defined by the $U_\text{c}$ and $V_k$ blocks. The total number of qubits is nine with four qubits dedicated to $N=16$ supporting points. For the Neumann boundaries, the even reflection of the register is prepared exclusively with Hadamard and CNOT gates, while the odd reflection of the Dirichlet condition also includes $Z$ gates.}
    \label{fig:lcu:twodim_PDE_BC}
\end{figure*}

The boundary method has several limitations. First, boundaries in each non-periodic direction must be of the same type and value, since the reflection depends on the boundary type, and different boundary values would produce discontinuities in the reflected field. Second, a prescribed non-zero derivative for inhomogeneous Neumann boundaries cannot be implemented using central second-order stencils on the reflected field without applying a correction source term to the PDE. These limitations on the boundaries are summarized for the heat equation in Tab.~\ref{tab:boundaries_limits}. 
\begin{table}[htpb]
\centering
\caption{Boundary conditions for Eqn.~\eqref{eq:heat_equation} supported by the approaches described in Sec.~\ref{sec:boundaries}.}
\label{tab:boundaries_limits}
\begin{tabular}{@{}lccc@{}}
            & Homogeneous   & Inhomogeneous \\
            \midrule
Dirichlet Reflection & \checkmark    &  \checkmark \\ 
Neumann Reflection   & \checkmark    &  $\times$ \\
Mixed Reflection     & \checkmark    &  $\times$ \\
            \midrule
Dirichlet no-reflection & $\times$     & $\times$ \\
Neumann no-reflection   & \checkmark   & $\times$
\end{tabular}
\end{table}

An alternative to the method of images for implementing homogeneous Neumann boundaries in the heat equation is presented next. In particular, the LCU method described in Sec.~\ref{sec:lcu} is combined with the unitary decomposition of the non-periodic time-marching operator from Eqn.~\eqref{eq:A_matrix}, incorporating the corresponding boundary modifications given in Tab.~\ref{tab:boundaries_LCU}, as 
\begin{equation}
    \frac{A_{\text{Neumann}}}{1-2r_\text{h}} = I + \frac{r_\text{h}}{1-2r_\text{h}}(S_1+S_2) \, , 
    \label{eq:A_Neu_decomp}
\end{equation}
with 
\begin{equation}
    S_1 =
    \begin{bmatrix}
        0      & 1 & \dots  &   &  0      \\
        1      & 0 &        &   &         \\
        \vdots &   & 0      & 1 &         \\
               &   & 1      & 0 &         \\
        0      &   & \dots  &   &  \ddots 
    \end{bmatrix} = I\otimes X \, , 
    \label{eq:S_1}
\end{equation}
and 
\begin{equation}
    S_2 =
    \begin{bmatrix}
        1 &   &   &   &   &         & 0 \\
          & 0 & 1 &   &   &         & \\
          & 1 & 0 &   &   &         & \\
          &   &   & 0 & 1 &         & \\
          &   &   & 1 & 0 &         & \\
          &   &   &    &   & \ddots & \\
        0 &   &   &    &   &        & 1
    \end{bmatrix} \,.
    \label{eq:S_2}
\end{equation}
\begin{table}[htpb]
\centering
\caption{Boundary term modifications in Eqn.~\eqref{eq:A_matrix}}
\label{tab:boundaries_LCU}
\begin{tabular}{@{}lcccc@{}}
            & $b_{1,1}$  & $b_{1,N}$ & $b_{N,1}$ & $b_{N,N}$ \\
            \midrule
Periodic    & $1{-}2r_\text{h}$ & $r_\text{h}$   & $r_\text{h}$ & $1{-}2r_\text{h}$ \\ 
Dirichlet   & $1{-}2r_\text{h}$ & 0       & 0     & $1{-}2r_\text{h}$ \\
Neumann     & $1{-}r_\text{h}$  & 0       & 0     & $1{-}r_\text{h}$
\end{tabular}
\end{table}

This direct embedding of the boundaries has the advantage of reducing the number of qubits required for the reflection, while preserving the optimal success probabilities per time step with linear time scaling \cite{Over2024b}. Analogously to periodic cases, the LCU block encoding of $A_{\text{Neumann}}$ results in $\alpha = 1$, which guarantees the convergence of the cumulative success probabilities for non-zero steady-state solutions, as shown in Sec.~\ref{sec:lcu}. 

\begin{figure}[ht]
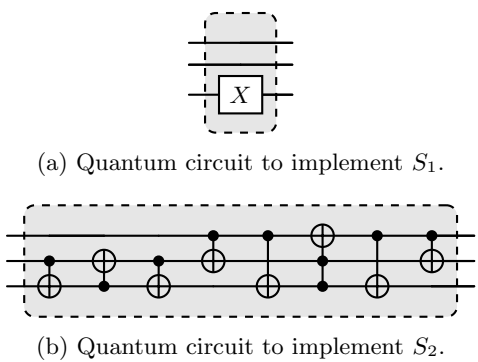

    \centering
    \begin{subfigure}[t]{0.49\textwidth}
        \centering
        \include{Figures/06-LCU-Diff_NoRefl_S1}
        \caption{Quantum circuit to implement $S_1$.}
        \label{fig:lcu:diff_S1}
    \end{subfigure}
    \begin{subfigure}[t]{0.49\textwidth}
        \centering
        \include{Figures/06-LCU-Diff_NoRefl_S2}
        \caption{Quantum circuit to implement $S_2$.}
        \label{fig:lcu:diff_S2}
    \end{subfigure}
    \caption{Three-qubit quantum circuits implementing $S_1$ and $S_2$ for homogeneous Neumann boundary conditions without reflection in the one-dimensional circuit of Fig.~\ref{fig:lcu:onedim}.}
    \label{fig:lcu:diff_noper_S12}
\end{figure}

The method can be extended to arbitrary dimensions. Here, one ($d=1$) and two ($d=2$) dimensions are described for the sake of clarity. For the one-dimensional case, the identity $I$ and the circuits of Fig.~\ref{fig:lcu:diff_noper_S12} for $S_1$ and $S_2$ are applied to the LCU circuit of Fig.~\ref{fig:lcu:onedim} with the coefficients $\kappa_0 = 1$, $\kappa_1 = \kappa_2 = r_\text{h}/(1-2\,d\,r_\text{h})$, and $\kappa_3=0$. For two dimensions, the five unitaries $I$, $I \otimes S_1$, $I \otimes S_2$, $S_1 \otimes I$, and $S_2 \otimes I$ are used in the LCU circuit of Fig.~\ref{fig:lcu:twodim} with the coefficients $\kappa_0 = 1$, $\kappa_1 = \kappa_2 =\kappa_3 = \kappa_4 = r_\text{h}/(1-2\,d\,r_\text{h})$, and $\kappa_5=\kappa_6=\kappa_7=0$. 

\section{Complexity}
\label{sec:complexity}
Given a problem in $d$ dimensions with $N=N^d_{\text{x}_1}$ equidistant supporting points, the spatial information is encoded in the quantum register using $log_2(N)$ qubits for the domain discretization, $log_2(3+2(d-1))$ ancilla qubits for the LCU method, and one additional qubit per non-periodic dimension independent of the type of boundary conditions. As a result, the total number of qubits scales as $O(log_2(N)+ log_2(d) + d)$. Furthermore, the number of attempts required for a successful execution of the quantum time-marching algorithm over the time horizon $T$ is intrinsically determined by the inverse of $p_\text{c}$ and it scales accordingly as $O(\|\phi(0)\|_2^2/\|\phi(T)\|_2^2)$. 

The temporal evolution is implemented by sequentially applying the quantum circuits in Fig.~\ref{fig:lcu:onedim_PDE_BC}, \ref{fig:lcu:twodim_PDE_BC}, or \ref{fig:lcu:twodim_PDE_MixedBC}. These quantum circuits consist of a constant number of gates per time step, and due to the absence of amplitude amplification, the gate complexity grows linearly with the number of time steps $N_\text{t} = T /\Delta t$. 
The limiting stability condition $r_\text{h}$ implies $\Delta t \propto \Delta x^2$, where $N_{\text{x}_1} \propto 1/\Delta x$, such that $N_\text{t}\propto TN_{\text{x}_1}^2$ and for $d$ dimensions $N_\text{t}\propto T N_{\text{x}_1}^{2} =  T N^{2/d}$. Given these contributions, the implementation has a gate complexity of $O((log_2(N)+ log_2(d) + d) T N^{2/d})$. 
The composition of the complexity is identical to that in Ref.~\cite{Over2024b} suppressing logarithmic terms, except for the treatment of boundary conditions, which implies doubling $N$ per dimension $d$. 

Analog to the quantum algorithm, the classical implementation also scales in time with $N_\text{t} \propto T{N_{\text{x}_1}^{2}}$ in one dimension. For $N=N_{\text{x}_1}^d$ supporting points distributed over $d$ dimensions, the classical complexity results instead in $O(N \, T N^{2/d})= O(TN^{(2+d)/d})$. Comparing the exponent of the complexities, the factor of $2/(2+d)$ indicates a polynomial speed up of the quantum algorithm over classical algorithms. As the approximation schemes in the quantum and classical implementations are identical, the underlying error of the Euler method and the second-order discretization is not taken into account.

\section{Simulations}
\label{sec:simulations}
The quantum methods for boundary implementation described above in Sec.~\ref{sec:boundaries} are investigated in this section using a state-vector simulation for the two-dimensional non-periodic heat equation (Eqn.~\eqref{eq:heat_equation}). Homogeneous Neumann, Dirichlet, and mixed boundaries are incorporated into the quantum time-marching framework with the method of images, while the direct embedding of the unitary decomposition in the time-marching operator is used to implement homogeneous Neumann conditions. For the latter, the matrix $A_\text{Neumann}$ is directly implemented with the LCU method to advance the linear transport by the quantum time-marching framework. Note that, for the evolution of $N_\text{t}$ time steps, the sequential execution of the circuits shown in Figs.~\ref{fig:lcu:onedim_PDE_BC}, \ref{fig:lcu:twodim_PDE_BC}, and \ref{fig:lcu:twodim_PDE_MixedBC} is equivalent to an $N_\text{t}$-sequence of $U_c$ gates, as the other quantum gates simplify to identities. 

The simulations refer to a spatial unit square $\vec{x} \in [0,1]\times[0,1]$ [m]. The initial condition ($t=0$) for all cases studied is based on a two-dimensional Gaussian perturbation of the form $\phi(\vec{x})=\exp\big(-200(x_1-0.25)^2 -200 (x_2-0.25)^2\big)$, placed in the lower left corner of the unit square. The temporal error measure $\varepsilon$ is defined by the l2-norm of the difference between classical FD and Quantum March (QM) results, i.e., $\varepsilon^t =\| \pmb{\phi}^t_\text{FD} - \pmb{\phi}^t_\text{QM}\|_2$. The probability of success $p^t$ and failure $1-p^t$ per time step, together with the cumulative probability of success $p_{\text{c}}$ per simulation time interval are also given in the results. 

\subsection{Neumann BC}
\label{sec:simulations:neumann}
This section reports the results obtained from two different implementations of homogeneous Neumann conditions along the route outlined by the methods described in Sec.~\ref{sec:boundaries}. 

\begin{figure*}[htpb]
    \centering
    \begin{subfigure}[t]{0.49\textwidth}
        \vspace{0pt}
        \includegraphics[width=\textwidth]{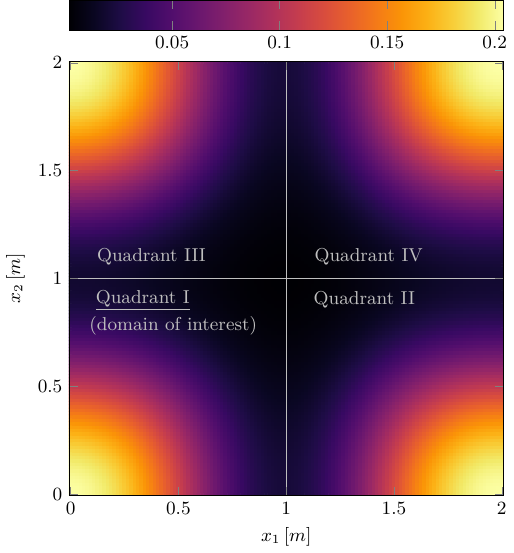}
        \caption{QM results of $\pmb{\phi}^t$ after $1000$ steps for the four reflected quadrants without the uncomputation of quadrants II, III, IV, and being quadrant I the domain of interest.}
        \label{fig:results:heat:Neumann:reflection:field}
    \end{subfigure}
    \begin{subfigure}[t]{0.49\textwidth}
        \vspace{0pt}
        \begin{subfigure}[t]{\textwidth}
            \includegraphics[width=\textwidth]{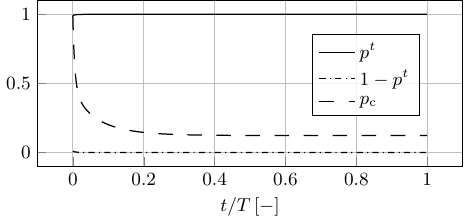}
            \caption{Temporal evolution of the algorithm probabilities in the state-vector simulation.}
            \label{fig:results:heat:Neumann:reflection:probs}
        \end{subfigure}
        \begin{subfigure}[t]{\textwidth}
            \includegraphics[width=\textwidth]{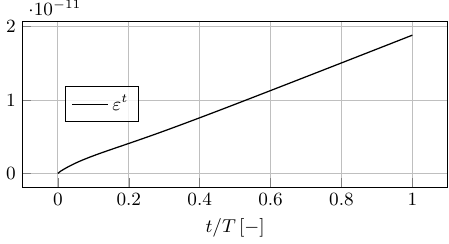}
            \caption{Temporal evolution of the error measure $\varepsilon^t$ w.r.t. classical FD results.}
            \label{fig:results:heat:Neumann:reflection:error}
        \end{subfigure}
    \end{subfigure}
    
    \caption{Comparison of QM and FD results of $\pmb{\phi}^t$ for the two-dimensional heat equation (Eqn.~\eqref{eq:heat_equation}) with homogeneous Neumann boundary conditions enforced by the method of images. Results refer to $128 \times 128$ grid points (including reflection quadrants) with $14$ working qubits and three ancilla qubits, obtained from state-vector simulation for $N_\text{t} = 12000$ time steps with $r_\text{h}=0.2$.}
    \label{fig:results:heat:Neumann:reflection}
\end{figure*}

First, the method of images is applied to enforce the boundary conditions using the reflection circuit illustrated in Fig.~\ref{fig:lcu:twodim_PDE_BC}. The $U_\text{c}$ gate of this circuit is defined by the controlled unitaries of the LCU circuit in Fig.~\ref{fig:lcu:twodim} to implement the time-marching operator $A$ of Eqn.~\eqref{eq:A_matrix}. To this end, we define $U_0 = I$, $U_1 = I \otimes S_0$, $U_2 = I \otimes S_0^\dagger$, $U_3 = S_0 \otimes I$, and $U_4 = S_0^\dagger \otimes I$, with the LCU coefficients as $\kappa_0 = 1$, $\kappa_1 = \kappa_2 =\kappa_3 = \kappa_4 = r_\text{h}/(1-2\,d\,r_\text{h})$, and $\kappa_5=\kappa_6=\kappa_7=0$. Furthermore, the quantum circuits given in Fig.~\ref{fig:lcu:diff_S0S0dagger} are used to implement the upper and lower diagonal shifts $S_0$ and $S_0^\dagger$. The results in Fig.~\ref{fig:results:heat:Neumann:reflection} depict the QM solution of $\pmb{\phi}^t$ computed for $N_\text{t}=12000$ explicit time steps with  $r_\text{h}=0.2$. The even (symmetric) reflection on the three additional quadrants II, III, and IV, without the register reset by the reflection part of the circuit in Fig.~\ref{fig:lcu:twodim_PDE_BC}, is indicated in Fig.~\ref{fig:results:heat:Neumann:reflection:field}, which corresponds to $1000$ time steps. The full resolution contains $128 \times 128$ grid points, where $N=64 \times 64$ points correspond to the (non-reflected) domain of interest in the first quadrant. A total of 17 qubits were used in these simulations, including 14 working qubits and three ancilla qubits. 

\begin{figure*}[htpb]
    \centering
    \begin{subfigure}[t]{0.49\textwidth}
        \hspace{0.8cm}
        \includegraphics{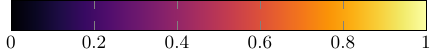}
        \centering
        \begin{subfigure}[t]{0.49\textwidth}
            \includegraphics{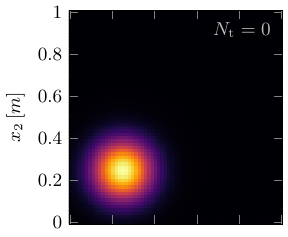}
            \label{fig:results:heat:Neumann_0}
        \end{subfigure}
        \hfill
        \begin{subfigure}[t]{0.49\textwidth}
            \hspace{0.1cm}
            \includegraphics{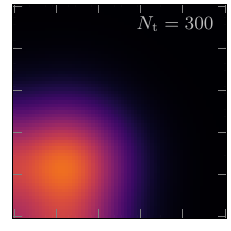}
            \label{fig:results:heat:Neumann_1}
        \end{subfigure}
        \begin{subfigure}[t]{0.49\textwidth}
            \includegraphics{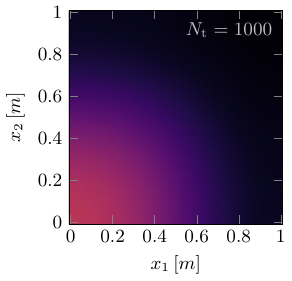}
            \label{fig:results:heat:Neumann_2}
        \end{subfigure}
        \hfill
        \begin{subfigure}[t]{0.49\textwidth}
            \hspace{0.121cm}
            \includegraphics{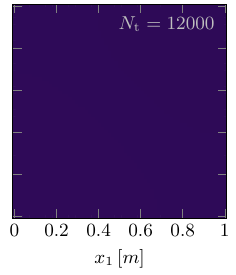}
            \label{fig:results:heat:Neumann_3}
        \end{subfigure}
        \caption{QM results of $\pmb{\phi}^t$ for $N_\text{t} = 0$, $300$, $1000$, and $12000$.}
        \label{fig:results:heat:Neumann:field}
    \end{subfigure}
    \hfill
    \begin{subfigure}[t]{0.49\textwidth}
        \vspace{0pt}
        \centering
        \begin{subfigure}[t]{\textwidth}
            \includegraphics{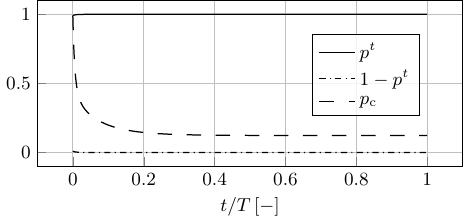}
            \caption{Temporal evolution of the algorithm probabilities in the state-vector simulation.} 
            \label{fig:results:heat:Neumann:probs}
        \end{subfigure}
        \begin{subfigure}[t]{\textwidth}
            \includegraphics{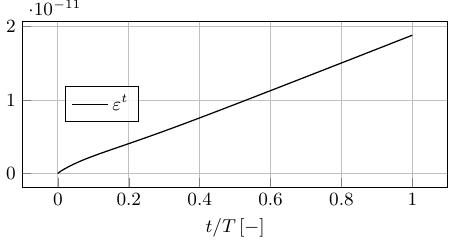}
            \caption{Temporal evolution of the error measure $\varepsilon^t$ w.r.t. classical FD results.}
            \label{fig:results:heat:Neumann:error}
        \end{subfigure}
    \end{subfigure}

    \caption{State-vector simulation of the two-dimensional heat equation (Eqn.~\eqref{eq:heat_equation}) with the homogeneous Neumann boundary conditions embedded directly in the time-marching operator. Results show $64 \times 64$ grid points for quadrant I (cf. Fig.~\ref{fig:results:heat:Neumann:reflection}) with $12$ working qubits and three ancilla qubits. The stability parameter is $r_\text{h}=0.2$ and $N_\text{t} = 12000$.}
    \label{fig:results:heat:Neumann}
\end{figure*}

Second, the homogeneous Neumann conditions are implemented without reflection by combining the controlled unitaries of the LCU circuit (Fig.~\ref{fig:lcu:twodim}) with the circuits given in Fig.~\ref{fig:lcu:diff_noper_S12}. To directly embed the two-dimensional matrix $A_\text{Neumann}$ of Eqn.~\eqref{eq:A_Neu_decomp}, the five unitaries in the LCU circuit are defined as $U_0 = I$, $U_1 = I \otimes S_1$, $U_2 = I \otimes S_2$, $U_3 =S_1 \otimes I$, and $U_4 =S_2 \otimes I$ with the same weights $\kappa_k$ as in the boundary reflection method. Figure~\ref{fig:results:heat:Neumann} displays the results of the QM simulations for an evolution of $\pmb{\phi}^t$ over $N_\text{t}=12000$ time steps, again setting $r_\text{h}=0.2$. Comparing the results of Fig.~\ref{fig:results:heat:Neumann:reflection:field} and Fig.~\ref{fig:results:heat:Neumann:field}, the different colorbars need to be taken into account. It should be noted that the absence of reflection quadrants spares two qubits, maintaining the resolution of $64 \times 64$ discrete points with $12$ working qubits and three ancilla qubits. 

The temporal evolution of the error $\varepsilon^t$ reaches maximum values of approximately $10^{-11}$ in Fig.~\ref{fig:results:heat:Neumann:reflection:error} and Fig.~\ref{fig:results:heat:Neumann:error}. This validation against classical FD unequivocally demonstrates the accuracy of the proposed QM boundary methods. 
Furthermore, the probability of success per time step converges to $p^t = \| \pmb{\phi}^{t+1} \|_2^2 / \| \pmb{\phi}^{t} \|_2^2 = 1$ as the steady state is approached because $\alpha = 1$ in the block encoding of the time-marching operator, cf. Sec.~\ref{sec:lcu}. The application of a non-unitary operation to represent the irreversible diffusion dynamics of the heat equation explains the decay in the cumulative probability of success $p_\text{c}$ for the transient solution, cf. Fig.~\ref{fig:results:heat:Neumann:reflection:probs} and Fig.~\ref{fig:results:heat:Neumann:probs}. However, the theoretically predicted $p_\text{c}$ in Sec.~\ref{sec:lcu} converges for both methods to $p_\text{c} = \| \pmb{\phi}^T \|_2^2/ \| \pmb{\phi}^0 \|_2^2 = 0.123$, cf. Fig.~\ref{fig:results:heat:Neumann:reflection:probs} and Fig.~\ref{fig:results:heat:Neumann:probs}, which is intrinsic to the problem. Although this final converged value of $p_\text{c}=0.123$ for the steady-state solution can only be increased at the cost of quadratic time complexity \cite{Fang2023}, the slope of its decay during the transient stage can be adjusted by modifying the time discretization, as $\Delta t$ determines the difference between the consecutive solutions $\pmb{\phi}^{t}$ and $\pmb{\phi}^{t+1}$. However, the selection of a smaller $\Delta t$ with an increase in the number of time steps will not affect the final convergence to the problem's intrinsic probabilities.

\subsection{Dirichlet BC}
\label{sec:simulations:dirichlet}
The homogeneous Dirichlet conditions are implemented by combining the quantum circuits of Fig.~\ref{fig:lcu:twodim} and Fig.~\ref{fig:lcu:twodim_PDE_BC}. Analogous to Sec.~\ref{sec:simulations:neumann}, the unitaries in $U_\text{c}$ building the time-marching operator $A$ of Eqn.~\eqref{eq:A_matrix} are selected as, $U_0=I$, $U_1=I \otimes S_0$, $U_2=I \otimes S_0^\dagger$, $U_3=S_0 \otimes I$, and $U_4=S_0^\dagger \otimes I$ with the circuits of Fig.~\ref{fig:lcu:diff_S0S0dagger} and the LCU weights $\kappa_0 = 1$, $\kappa_1 = \kappa_2 =\kappa_3 = \kappa_4 = r_\text{h}/(1-2\,d\,r_\text{h})$, $\kappa_5=\kappa_6=\kappa_7=0$ with $d=2$. The approach applied here differs from the one described in Sec.~\ref{sec:simulations:neumann} only in the type of reflection. The method of images maps the initial conditions by applying an odd (antisymmetric) reflection. This is achieved using the $Z$-gates in Fig.~\ref{fig:lcu:twodim_PDE_BC}. 
The full resolution is $128 \times 128$ with $N=64 \times 64$ grid points dedicated to the quadrant of interest (quadrant I). Again, a total of $17$ qubits including three ancilla qubits are used in the state-vector simulation of $N_\text{t} = 12000$ time steps in combination with $r_\text{h} = 0.2$. 

\begin{figure*}[htpb]
    \centering
    \begin{subfigure}[t]{0.49\textwidth}
        \hspace{0.8cm}
        \includegraphics{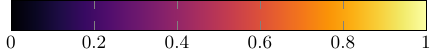}
        \centering
        \begin{subfigure}[t]{0.49\textwidth}
            \includegraphics{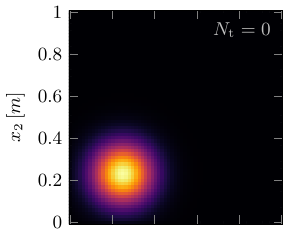}
            \label{fig:results:heat:Dirichlet_0}
        \end{subfigure}
        \hfill
        \begin{subfigure}[t]{0.49\textwidth}
            \hspace{0.1cm}
            \includegraphics{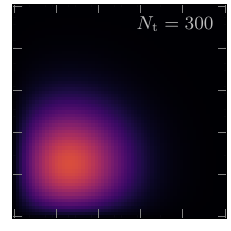}
            \label{fig:results:heat:Dirichlet_1}
        \end{subfigure}
        \begin{subfigure}[t]{0.49\textwidth}
            \includegraphics{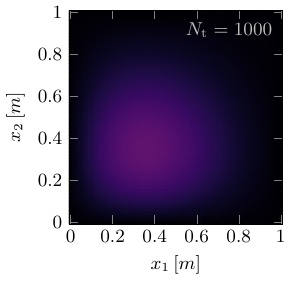}
            \label{fig:results:heat:Dirichlet_2}
        \end{subfigure}
        \hfill
        \begin{subfigure}[t]{0.49\textwidth}
            \hspace{0.121cm}
            \includegraphics{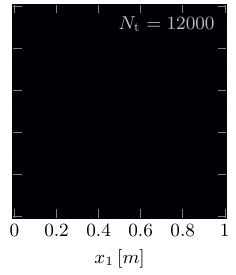}
            \label{fig:results:heat:Dirichlet_3}
        \end{subfigure}
        \caption{QM results of $\pmb{\phi}^t$ for $N_\text{t} = 0$, $300$, $1000$, and $12000$.}
        \label{fig:results:heat:Dirichlet:field}
    \end{subfigure}
    \hfill
    \begin{subfigure}[t]{0.49\textwidth}
        \vspace{0pt}
        \centering
        \begin{subfigure}[t]{\textwidth}
            \includegraphics{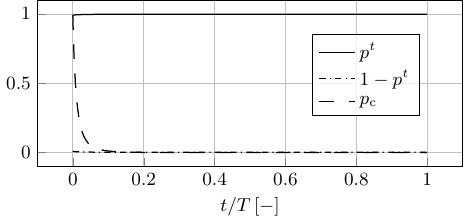}
            \caption{Temporal evolution of the algorithm probabilities in the state-vector simulation.}
            \label{fig:results:heat:Dirichlet:probs}
        \end{subfigure}
        \begin{subfigure}[t]{\textwidth}
            \includegraphics{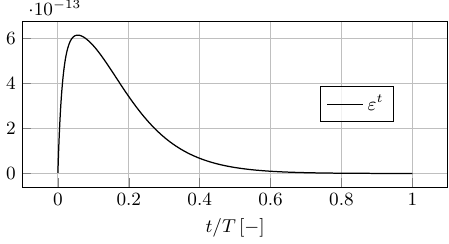}
            \caption{Temporal evolution of the error measure $\varepsilon^t$ w.r.t. classical FD results. Convergence to zero due to a vanishing steady solution.}
            \label{fig:results:heat:Dirichlet:error}
        \end{subfigure}
    \end{subfigure}

    \caption{Benchmarking of QM results for the two-dimensional heat equation (Eqn.~\eqref{eq:heat_equation}) with homogeneous Dirichlet boundary conditions enforced by the method of images. Results refer to $128 \times 128$ grid points (including reflection quadrants) with $14$ working qubits and three ancilla qubits. State-vector simulation for $N_\text{t} = 12000$ time steps with $r_\text{h}=0.2$.}
    \label{fig:results:heat:Dirichlet}
\end{figure*}

The results are depicted in Fig.~\ref{fig:results:heat:Dirichlet}. This figure shows $\pmb{\phi}^t$ for four exemplary time instants ($N_\text{t}=0,300,1000,12000$) of the QM temporal evolution in quadrant I (Fig.~\ref{fig:results:heat:Dirichlet:field}), the algorithm probabilities (Fig.~\ref{fig:results:heat:Dirichlet:probs}), and the error measure $\varepsilon^t$ over time (Fig.~\ref{fig:results:heat:Dirichlet:error}). 
The validation of the method is clearly proven by the resulting low error values, with maximums of the order of $10^{-13}$. 
As shown in Fig.~\ref{fig:results:heat:Dirichlet:probs}, the probability of success per time step $p^t$ converges to one. 
However, the cumulative probability $p_\text{c}$ vanishes as the steady-state solution is reached. 
This is due to the homogeneous Dirichlet boundaries that force the steady-state solution to converge to zero, and consequently, the numerator in $p_\text{c} = \| \pmb{\phi}^T \|_2^2/ \| \pmb{\phi}^0 \|_2^2$ also vanishes. 
Introducing inhomogeneous Dirichlet conditions would resolve this issue. 
Nevertheless, such an approach requires shifting the entire quantum register by a constant value, which is challenging to implement with universal quantum gates. 
To address this difficulty, TT-based techniques \cite{vanHuelst2025} such as Tensor Quantum Programming (TQP) \cite{Termanova2024,Siegl2025} can be employed to initialize the register, encoding the inhomogeneity into quantum gates. 

\subsection{Mixed BC}
\label{sec:simulations:mixed}
Finally, homogeneous mixed boundaries are simulated with the method described in Sec.~\ref{sec:boundaries}. In the configuration under investigation, Neumann conditions are applied to the exterior boundaries whose normal points in the $x_1$ direction and Dirichlet conditions are applied in the $x_2$ direction, which implies an even (symmetric) and an odd (antisymmetric) reflection of the initial conditions, respectively. 

For the implementation, the reflection circuit of Fig.~\ref{fig:lcu:twodim_PDE_MixedBC} is combined with the unitary block $U_\text{c}$ of the LCU circuit in Fig.~\ref{fig:lcu:twodim}. The two-dimensional time-marching operator $A$ of Eqn.~\eqref{eq:A_matrix} is constructed by means of $U_0=I$, $U_1=I \otimes S_0$, $U_2=I \otimes S_0^\dagger$, $U_3=S_0 \otimes I$, and $U_4=S_0^\dagger \otimes I$. To this end, the circuits of Fig.~\ref{fig:lcu:diff_S0S0dagger} and the LCU coefficients $\kappa_0 = 1$, $\kappa_1 = \kappa_2 =\kappa_3 = \kappa_4 = r_\text{h}/(1-2\,d\,r_\text{h})$, and $\kappa_5=\kappa_6=\kappa_7=0$ with $d=2$ are employed. The simulation covers $N_\text{t}=12000$ time steps with the stability parameter $r_\text{h} = 0.2$. The total spatial resolution is defined as $128 \times 128$ supporting points, including the reflection quadrants II, III, IV, where $N=64 \times 64$ points are used for the quadrant of interest (I), cf. Fig.~\ref{fig:results:heat:Neumann:reflection}. The employed $17$ qubits are divided into $14$ working qubits and three ancilla qubits. 

\begin{figure*}[htpb]
    \centering
    \begin{subfigure}[t]{0.49\textwidth}
        \hspace{0.8cm}
        \includegraphics{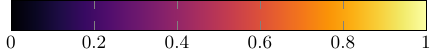}
        \centering
        \begin{subfigure}[t]{0.49\textwidth}
            \includegraphics{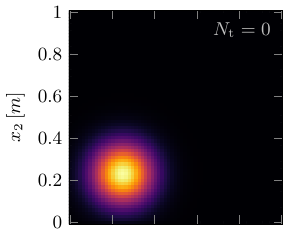}
            \label{fig:results:heat:Mixed_0}
        \end{subfigure}
        \hfill
        \begin{subfigure}[t]{0.49\textwidth}
            \hspace{0.1cm}
            \includegraphics{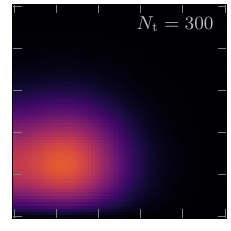}
            \label{fig:results:heat:Mixed_1}
        \end{subfigure}
        \begin{subfigure}[t]{0.49\textwidth}
            \includegraphics{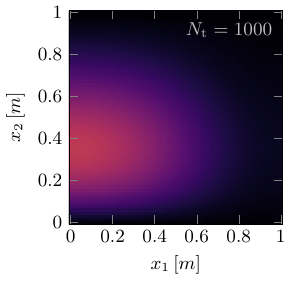}
            \label{fig:results:heat:Mixed_2}
        \end{subfigure}
        \hfill
        \begin{subfigure}[t]{0.49\textwidth}
            \hspace{0.121cm}
            \includegraphics{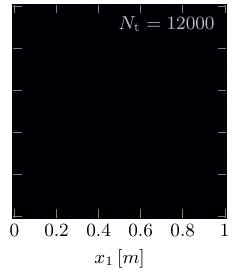}
            \label{fig:results:heat:Mixed_3}
        \end{subfigure}
        \caption{QM results of $\pmb{\phi}^t$ for $N_\text{t} = 0$, $300$, $1000$, and $12000$.}
        \label{fig:results:heat:Mixed:field}
    \end{subfigure}
    \hfill
    \begin{subfigure}[t]{0.49\textwidth}
        \vspace{0pt}
        \centering
        \begin{subfigure}[t]{\textwidth}
            \includegraphics{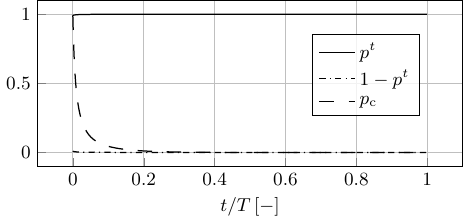}
            \caption{Temporal evolution of the algorithm probabilities in the state-vector simulation.}
            \label{fig:results:heat:Mixed:probs}
        \end{subfigure}
        \begin{subfigure}[t]{\textwidth}
            \includegraphics{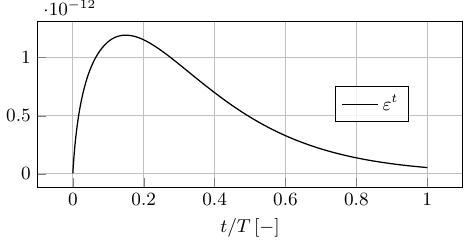}
            \caption{Temporal evolution of the error measure $\varepsilon^t$ w.r.t. classical FD results. Convergence to zero due to a vanishing steady solution.}
            \label{fig:results:heat:Mixed:error}
        \end{subfigure}
    \end{subfigure}

    \caption{QM results for the two-dimensional heat equation (Eqn.~\eqref{eq:heat_equation}) with homogeneous mixed boundary conditions enforced by the method of images. The horizontal direction corresponds to Neumann while the vertical direction to Dirichlet. Results refer to $128 \times 128$ grid points (including reflection quadrants) with $14$ working qubits and three ancilla qubits. State-vector simulation for $N_\text{t} = 12000$ time steps with $r_\text{h}=0.2$.}
    \label{fig:results:heat:Mixed}
\end{figure*}

Figure~\ref{fig:results:heat:Mixed:field} depicts the temporal evolution of $\pmb{\phi}^t$ in quadrant I for four exemplary time instants, i.e., $N_\text{t}=0,300,1000,12000$. There, the different effects of the boundaries on the spatial distribution of $\pmb{\phi}^t$ can clearly be observed. As in the previous experiments, the error values in Fig.~\ref{fig:results:heat:Mixed:error} show excellent agreement with results obtained from classical FD approaches, with maximum values in the order of $10^{-12}$. 

Paying attention to the probabilities of success, Fig.~\ref{fig:results:heat:Mixed:probs} illustrates the convergence of $p^t$ to unity and $p_\text{c}$ to zero. The latter is a consequence of the steady-state solution approaching zero due to the homogeneous Dirichlet condition along the $x_2$ boundaries. As discussed in Sec.~\ref{sec:simulations:dirichlet}, this limitation can be addressed by solving for non-homogeneous boundary conditions. To this end, the problem is substituted using a smooth extension, which requires adding a non-zero constant to the initial and the boundary conditions, to recover the homogeneous solution. The construction of the corresponding quantum gates to encode this adjustment can be implemented by TT techniques \cite{Termanova2024}. 

\section{Conclusion}
\label{sec:conclusions}
This paper presents the complete implementation of time-marching algorithms to solve diffusion problems described by second-order PDEs with different types of boundary conditions on fault-tolerant QCs. The proposed approaches combine the LCU algorithm to capture the diffusive dynamics and the method of images to enforce boundaries. 
In particular, the implementation of diffusion while maintaining optimal success probabilities constitutes the strongest asset of the method. Moreover, it is noteworthy to remark that the logarithmic scaling in the number of spatial points makes the two-fold increase in the reflection for the boundary treatment particularly cheap and efficient on a QC. In the specific case of Neumann boundaries, the domain reflection in each non-periodic direction is replaced by directly block encoding the boundary into the unitary decomposition of the time-marching operator. 

The presented results show unequivocally that all types of boundaries can be implemented on a fault-tolerant QC with very high accuracy compared to classical Euler-FD discretizations. Furthermore, optimal success probabilities with linear time scaling are demonstrated for two-dimensional simulations, while the proposed approach can straightforwardly be extended to arbitrary dimensions. The treatment of diffusive dynamics shows potential for implementing other non-unitary processes, while an application of the boundary treatment to other initial-boundary value problems in structural, fluid, or quantum mechanics is an interesting direction of future research.

\begin{acknowledgments}
This work and the research are supported by the \href{https://doi.org/10.3030/101080085}{European Union's Horizon Europe research and innovation program (HORIZON-CL4-2021-DIGITAL-EMERGING-02-10) under grant agreement No. 101080085 \textsc{QCFD}}. Furthermore, we thank Peter Brearley for the fruitful discussions.
\end{acknowledgments}

\section*{Author Contributions}
\addcontentsline{toc}{section}{Author Contributions}
\textbf{Sergio~Bengoechea:} Conceptualization, Formal analysis, Investigation, Methodology, Software, Validation, Visualization, Writing - original draft, Writing - review \& editing. 
\textbf{Paul~Over:} Conceptualization, Formal analysis, Investigation, Methodology, Software, Validation, Visualization, Writing - original draft, Writing - review \& editing. 
\textbf{Thomas~Rung:} Funding acquisition, Project administration, Supervision, Resources,  Writing - original draft, Writing - review \& editing. 

\section*{Data availability}
\addcontentsline{toc}{section}{Data availability}
The data is available via \href{https://doi.org/10.25592/uhhfdm.18048}{doi.org/10.25592/uhhfdm.18048}.

\bibliography{apssamp}

\begin{thebibliography}{52}%
\makeatletter
\providecommand \@ifxundefined [1]{%
 \@ifx{#1\undefined}
}%
\providecommand \@ifnum [1]{%
 \ifnum #1\expandafter \@firstoftwo
 \else \expandafter \@secondoftwo
 \fi
}%
\providecommand \@ifx [1]{%
 \ifx #1\expandafter \@firstoftwo
 \else \expandafter \@secondoftwo
 \fi
}%
\providecommand \natexlab [1]{#1}%
\providecommand \enquote  [1]{``#1''}%
\providecommand \bibnamefont  [1]{#1}%
\providecommand \bibfnamefont [1]{#1}%
\providecommand \citenamefont [1]{#1}%
\providecommand \href@noop [0]{\@secondoftwo}%
\providecommand \href [0]{\begingroup \@sanitize@url \@href}%
\providecommand \@href[1]{\@@startlink{#1}\@@href}%
\providecommand \@@href[1]{\endgroup#1\@@endlink}%
\providecommand \@sanitize@url [0]{\catcode `\\12\catcode `\$12\catcode
  `\&12\catcode `\#12\catcode `\^12\catcode `\_12\catcode `\%12\relax}%
\providecommand \@@startlink[1]{}%
\providecommand \@@endlink[0]{}%
\providecommand \url  [0]{\begingroup\@sanitize@url \@url }%
\providecommand \@url [1]{\endgroup\@href {#1}{\urlprefix }}%
\providecommand \urlprefix  [0]{URL }%
\providecommand \Eprint [0]{\href }%
\providecommand \doibase [0]{https://doi.org/}%
\providecommand \selectlanguage [0]{\@gobble}%
\providecommand \bibinfo  [0]{\@secondoftwo}%
\providecommand \bibfield  [0]{\@secondoftwo}%
\providecommand \translation [1]{[#1]}%
\providecommand \BibitemOpen [0]{}%
\providecommand \bibitemStop [0]{}%
\providecommand \bibitemNoStop [0]{.\EOS\space}%
\providecommand \EOS [0]{\spacefactor3000\relax}%
\providecommand \BibitemShut  [1]{\csname bibitem#1\endcsname}%
\let\auto@bib@innerbib\@empty
\bibitem [{\citenamefont {{Aslam Bhutta}}\ \emph {et~al.}(2012)\citenamefont
  {{Aslam Bhutta}}, \citenamefont {Hayat}, \citenamefont {Bashir},
  \citenamefont {Khan}, \citenamefont {Ahmad},\ and\ \citenamefont
  {Khan}}]{Bhutta2012}%
  \BibitemOpen
  \bibfield  {author} {\bibinfo {author} {\bibfnamefont {M.~M.}\ \bibnamefont
  {{Aslam Bhutta}}}, \bibinfo {author} {\bibfnamefont {N.}~\bibnamefont
  {Hayat}}, \bibinfo {author} {\bibfnamefont {M.~H.}\ \bibnamefont {Bashir}},
  \bibinfo {author} {\bibfnamefont {A.~R.}\ \bibnamefont {Khan}}, \bibinfo
  {author} {\bibfnamefont {K.~N.}\ \bibnamefont {Ahmad}},\ and\ \bibinfo
  {author} {\bibfnamefont {S.}~\bibnamefont {Khan}},\ }\href
  {https://doi.org/https://doi.org/10.1016/j.applthermaleng.2011.09.001}
  {\bibfield  {journal} {\bibinfo  {journal} {Applied Thermal Engineering}\
  }\textbf {\bibinfo {volume} {32}},\ \bibinfo {pages} {1} (\bibinfo {year}
  {2012})}\BibitemShut {NoStop}%
\bibitem [{\citenamefont {Jameson}(1988)}]{Jameson1988}%
  \BibitemOpen
  \bibfield  {author} {\bibinfo {author} {\bibfnamefont {A.}~\bibnamefont
  {Jameson}},\ }\href {https://doi.org/10.1007/bf01061285} {\bibfield
  {journal} {\bibinfo  {journal} {Journal of Scientific Computing}\ }\textbf
  {\bibinfo {volume} {3}},\ \bibinfo {pages} {233} (\bibinfo {year}
  {1988})}\BibitemShut {NoStop}%
\bibitem [{\citenamefont {Zhang}\ \emph {et~al.}(2021)\citenamefont {Zhang},
  \citenamefont {Wang},\ and\ \citenamefont {Yan}}]{Zhang2021}%
  \BibitemOpen
  \bibfield  {author} {\bibinfo {author} {\bibfnamefont {Z.}~\bibnamefont
  {Zhang}}, \bibinfo {author} {\bibfnamefont {X.}~\bibnamefont {Wang}},\ and\
  \bibinfo {author} {\bibfnamefont {Y.}~\bibnamefont {Yan}},\ }\href
  {https://doi.org/https://doi.org/10.1016/j.prime.2021.100009} {\bibfield
  {journal} {\bibinfo  {journal} {e-Prime - Advances in Electrical Engineering,
  Electronics and Energy}\ }\textbf {\bibinfo {volume} {1}},\ \bibinfo {pages}
  {100009} (\bibinfo {year} {2021})}\BibitemShut {NoStop}%
\bibitem [{\citenamefont {Vallis}(2017)}]{Vallis2017}%
  \BibitemOpen
  \bibfield  {author} {\bibinfo {author} {\bibfnamefont {G.~K.}\ \bibnamefont
  {Vallis}},\ }\href {https://doi.org/https://doi.org/10.1017/9781107588417}
  {\emph {\bibinfo {title} {Atmospheric and Oceanic Fluid Dynamics:
  Fundamentals and Large-Scale Circulation}}},\ \bibinfo {edition} {2nd}\ ed.\
  (\bibinfo  {publisher} {Cambridge University Press},\ \bibinfo {year}
  {2017})\BibitemShut {NoStop}%
\bibitem [{\citenamefont {Khan}\ \emph {et~al.}(2018)\citenamefont {Khan},
  \citenamefont {Hounshell},\ and\ \citenamefont {Fuchs}}]{Khan2018}%
  \BibitemOpen
  \bibfield  {author} {\bibinfo {author} {\bibfnamefont {H.~N.}\ \bibnamefont
  {Khan}}, \bibinfo {author} {\bibfnamefont {D.~A.}\ \bibnamefont
  {Hounshell}},\ and\ \bibinfo {author} {\bibfnamefont {E.~R.~H.}\ \bibnamefont
  {Fuchs}},\ }\href {https://doi.org/10.1038/s41928-017-0005-9} {\bibfield
  {journal} {\bibinfo  {journal} {Nature Electronics}\ }\textbf {\bibinfo
  {volume} {1}},\ \bibinfo {pages} {14} (\bibinfo {year} {2018})}\BibitemShut
  {NoStop}%
\bibitem [{\citenamefont {Esmaeilzadeh}\ \emph {et~al.}(2011)\citenamefont
  {Esmaeilzadeh}, \citenamefont {Blem}, \citenamefont {Amant}, \citenamefont
  {Sankaralingam},\ and\ \citenamefont {Burger}}]{Esmaeilzadeh2011}%
  \BibitemOpen
  \bibfield  {author} {\bibinfo {author} {\bibfnamefont {H.}~\bibnamefont
  {Esmaeilzadeh}}, \bibinfo {author} {\bibfnamefont {E.}~\bibnamefont {Blem}},
  \bibinfo {author} {\bibfnamefont {R.~S.}\ \bibnamefont {Amant}}, \bibinfo
  {author} {\bibfnamefont {K.}~\bibnamefont {Sankaralingam}},\ and\ \bibinfo
  {author} {\bibfnamefont {D.}~\bibnamefont {Burger}},\ }in\ \href
  {https://ieeexplore.ieee.org/abstract/document/6307773} {\emph {\bibinfo
  {booktitle} {2011 38th Annual International Symposium on Computer
  Architecture (ISCA)}}}\ (\bibinfo {year} {2011})\ pp.\ \bibinfo {pages}
  {365--376}\BibitemShut {NoStop}%
\bibitem [{\citenamefont {Slotnick}\ \emph {et~al.}(2014)\citenamefont
  {Slotnick}, \citenamefont {Khodadoust}, \citenamefont {Alonso}, \citenamefont
  {Darmofal}, \citenamefont {Gropp}, \citenamefont {Lurie},\ and\ \citenamefont
  {Mavriplis}}]{Nasa2030}%
  \BibitemOpen
  \bibfield  {author} {\bibinfo {author} {\bibfnamefont {J.~P.}\ \bibnamefont
  {Slotnick}}, \bibinfo {author} {\bibfnamefont {A.}~\bibnamefont
  {Khodadoust}}, \bibinfo {author} {\bibfnamefont {J.}~\bibnamefont {Alonso}},
  \bibinfo {author} {\bibfnamefont {D.}~\bibnamefont {Darmofal}}, \bibinfo
  {author} {\bibfnamefont {W.}~\bibnamefont {Gropp}}, \bibinfo {author}
  {\bibfnamefont {E.}~\bibnamefont {Lurie}},\ and\ \bibinfo {author}
  {\bibfnamefont {D.~J.}\ \bibnamefont {Mavriplis}},\ }\href
  {https://ntrs.nasa.gov/citations/20140003093} {\emph {\bibinfo {title} {CFD
  vision 2030 study: a path to revolutionary computational aerosciences}}},\
  \bibinfo {type} {Tech. Rep.}\ (\bibinfo  {institution} {NASA},\ \bibinfo
  {year} {2014})\BibitemShut {NoStop}%
\bibitem [{\citenamefont {Jaksch}\ \emph {et~al.}(2023)\citenamefont {Jaksch},
  \citenamefont {Givi}, \citenamefont {Daley},\ and\ \citenamefont
  {Rung}}]{Jaksch2023}%
  \BibitemOpen
  \bibfield  {author} {\bibinfo {author} {\bibfnamefont {D.}~\bibnamefont
  {Jaksch}}, \bibinfo {author} {\bibfnamefont {P.}~\bibnamefont {Givi}},
  \bibinfo {author} {\bibfnamefont {A.~J.}\ \bibnamefont {Daley}},\ and\
  \bibinfo {author} {\bibfnamefont {T.}~\bibnamefont {Rung}},\ }\href
  {https://doi.org/10.2514/1.J062426} {\bibfield  {journal} {\bibinfo
  {journal} {AIAA Journal}\ }\textbf {\bibinfo {volume} {61}},\ \bibinfo
  {pages} {1885} (\bibinfo {year} {2023})}\BibitemShut {NoStop}%
\bibitem [{\citenamefont {Givi}\ \emph {et~al.}(2020)\citenamefont {Givi},
  \citenamefont {Daley}, \citenamefont {Mavriplis},\ and\ \citenamefont
  {Malik}}]{Givi2020}%
  \BibitemOpen
  \bibfield  {author} {\bibinfo {author} {\bibfnamefont {P.}~\bibnamefont
  {Givi}}, \bibinfo {author} {\bibfnamefont {A.~J.}\ \bibnamefont {Daley}},
  \bibinfo {author} {\bibfnamefont {D.}~\bibnamefont {Mavriplis}},\ and\
  \bibinfo {author} {\bibfnamefont {M.}~\bibnamefont {Malik}},\ }\href
  {https://doi.org/10.2514/1.J059183} {\bibfield  {journal} {\bibinfo
  {journal} {AIAA Journal}\ }\textbf {\bibinfo {volume} {58}},\ \bibinfo
  {pages} {3715} (\bibinfo {year} {2020})}\BibitemShut {NoStop}%
\bibitem [{\citenamefont {Nielsen}\ and\ \citenamefont
  {Chuang}(2010)}]{Nielsen2010}%
  \BibitemOpen
  \bibfield  {author} {\bibinfo {author} {\bibfnamefont {M.~A.}\ \bibnamefont
  {Nielsen}}\ and\ \bibinfo {author} {\bibfnamefont {I.~L.}\ \bibnamefont
  {Chuang}},\ }\href {https://doi.org/10.1017/CBO9780511976667} {\emph
  {\bibinfo {title} {Quantum Computation and Quantum Information}}}\ (\bibinfo
  {publisher} {Cambridge University Press},\ \bibinfo {year}
  {2010})\BibitemShut {NoStop}%
\bibitem [{\citenamefont {Preskill}(2018)}]{Preskill2018}%
  \BibitemOpen
  \bibfield  {author} {\bibinfo {author} {\bibfnamefont {J.}~\bibnamefont
  {Preskill}},\ }\href {https://doi.org/10.22331/q-2018-08-06-79} {\bibfield
  {journal} {\bibinfo  {journal} {{Quantum}}\ }\textbf {\bibinfo {volume}
  {2}},\ \bibinfo {pages} {79} (\bibinfo {year} {2018})}\BibitemShut {NoStop}%
\bibitem [{\citenamefont {Cerezo}\ \emph {et~al.}(2021)\citenamefont {Cerezo},
  \citenamefont {Arrasmith}, \citenamefont {Babbush}, \citenamefont {Benjamin},
  \citenamefont {Endo}, \citenamefont {Fujii}, \citenamefont {McClean},
  \citenamefont {Mitarai}, \citenamefont {Yuan}, \citenamefont {Cincio},\ and\
  \citenamefont {Coles}}]{Cerezo2021}%
  \BibitemOpen
  \bibfield  {author} {\bibinfo {author} {\bibfnamefont {M.}~\bibnamefont
  {Cerezo}}, \bibinfo {author} {\bibfnamefont {A.}~\bibnamefont {Arrasmith}},
  \bibinfo {author} {\bibfnamefont {R.}~\bibnamefont {Babbush}}, \bibinfo
  {author} {\bibfnamefont {S.~C.}\ \bibnamefont {Benjamin}}, \bibinfo {author}
  {\bibfnamefont {S.}~\bibnamefont {Endo}}, \bibinfo {author} {\bibfnamefont
  {K.}~\bibnamefont {Fujii}}, \bibinfo {author} {\bibfnamefont {J.~R.}\
  \bibnamefont {McClean}}, \bibinfo {author} {\bibfnamefont {K.}~\bibnamefont
  {Mitarai}}, \bibinfo {author} {\bibfnamefont {X.}~\bibnamefont {Yuan}},
  \bibinfo {author} {\bibfnamefont {L.}~\bibnamefont {Cincio}},\ and\ \bibinfo
  {author} {\bibfnamefont {P.~J.}\ \bibnamefont {Coles}},\ }\href
  {https://doi.org/10.1038/s42254-021-00348-9} {\bibfield  {journal} {\bibinfo
  {journal} {Nature Reviews Physics}\ }\textbf {\bibinfo {volume} {3}},\
  \bibinfo {pages} {2522} (\bibinfo {year} {2021})}\BibitemShut {NoStop}%
\bibitem [{\citenamefont {Lubasch}\ \emph {et~al.}(2020)\citenamefont
  {Lubasch}, \citenamefont {Joo}, \citenamefont {Moinier}, \citenamefont
  {Kiffner},\ and\ \citenamefont {Jaksch}}]{Lubasch2020}%
  \BibitemOpen
  \bibfield  {author} {\bibinfo {author} {\bibfnamefont {M.}~\bibnamefont
  {Lubasch}}, \bibinfo {author} {\bibfnamefont {J.}~\bibnamefont {Joo}},
  \bibinfo {author} {\bibfnamefont {P.}~\bibnamefont {Moinier}}, \bibinfo
  {author} {\bibfnamefont {M.}~\bibnamefont {Kiffner}},\ and\ \bibinfo {author}
  {\bibfnamefont {D.}~\bibnamefont {Jaksch}},\ }\href
  {https://link.aps.org/doi/10.1103/PhysRevA.101.010301} {\bibfield  {journal}
  {\bibinfo  {journal} {Physical Review A}\ }\textbf {\bibinfo {volume} {101}}
  (\bibinfo {year} {2020})}\BibitemShut {NoStop}%
\bibitem [{\citenamefont {Over}\ \emph
  {et~al.}(2025{\natexlab{a}})\citenamefont {Over}, \citenamefont {Bengoechea},
  \citenamefont {Rung}, \citenamefont {Clerici}, \citenamefont {Scandurra},
  \citenamefont {{de Villiers}},\ and\ \citenamefont {Jaksch}}]{Over2024a}%
  \BibitemOpen
  \bibfield  {author} {\bibinfo {author} {\bibfnamefont {P.}~\bibnamefont
  {Over}}, \bibinfo {author} {\bibfnamefont {S.}~\bibnamefont {Bengoechea}},
  \bibinfo {author} {\bibfnamefont {T.}~\bibnamefont {Rung}}, \bibinfo {author}
  {\bibfnamefont {F.}~\bibnamefont {Clerici}}, \bibinfo {author} {\bibfnamefont
  {L.}~\bibnamefont {Scandurra}}, \bibinfo {author} {\bibfnamefont
  {E.}~\bibnamefont {{de Villiers}}},\ and\ \bibinfo {author} {\bibfnamefont
  {D.}~\bibnamefont {Jaksch}},\ }\href
  {https://doi.org/10.1016/j.compfluid.2024.106508} {\bibfield  {journal}
  {\bibinfo  {journal} {Computers \& Fluids}\ }\textbf {\bibinfo {volume}
  {288}},\ \bibinfo {pages} {106508} (\bibinfo {year}
  {2025}{\natexlab{a}})}\BibitemShut {NoStop}%
\bibitem [{\citenamefont {Bengoechea}\ \emph {et~al.}(2025)\citenamefont
  {Bengoechea}, \citenamefont {Over}, \citenamefont {Jaksch},\ and\
  \citenamefont {Rung}}]{Bengoechea2025}%
  \BibitemOpen
  \bibfield  {author} {\bibinfo {author} {\bibfnamefont {S.}~\bibnamefont
  {Bengoechea}}, \bibinfo {author} {\bibfnamefont {P.}~\bibnamefont {Over}},
  \bibinfo {author} {\bibfnamefont {D.}~\bibnamefont {Jaksch}},\ and\ \bibinfo
  {author} {\bibfnamefont {T.}~\bibnamefont {Rung}},\ }\href
  {https://doi.org/10.2514/1.J065582} {\bibfield  {journal} {\bibinfo
  {journal} {AIAA Journal}\ }\textbf {\bibinfo {volume} {(in-press)}} (\bibinfo
  {year} {2025})}\BibitemShut {NoStop}%
\bibitem [{\citenamefont {Sato}\ \emph {et~al.}(2024)\citenamefont {Sato},
  \citenamefont {Kondo}, \citenamefont {Hamamura}, \citenamefont {Onodera},\
  and\ \citenamefont {Yamamoto}}]{Sato2024}%
  \BibitemOpen
  \bibfield  {author} {\bibinfo {author} {\bibfnamefont {Y.}~\bibnamefont
  {Sato}}, \bibinfo {author} {\bibfnamefont {R.}~\bibnamefont {Kondo}},
  \bibinfo {author} {\bibfnamefont {I.}~\bibnamefont {Hamamura}}, \bibinfo
  {author} {\bibfnamefont {T.}~\bibnamefont {Onodera}},\ and\ \bibinfo {author}
  {\bibfnamefont {N.}~\bibnamefont {Yamamoto}},\ }\href
  {https://link.aps.org/doi/10.1103/PhysRevResearch.6.033246} {\bibfield
  {journal} {\bibinfo  {journal} {Phys. Rev. Res.}\ }\textbf {\bibinfo {volume}
  {6}},\ \bibinfo {pages} {033246} (\bibinfo {year} {2024})}\BibitemShut
  {NoStop}%
\bibitem [{\citenamefont {Alipanah}\ \emph {et~al.}(2025)\citenamefont
  {Alipanah}, \citenamefont {Zhang}, \citenamefont {Yao}, \citenamefont
  {Thompson}, \citenamefont {Nguyen}, \citenamefont {Liu}, \citenamefont
  {Givi}, \citenamefont {McDermott},\ and\ \citenamefont
  {Mendoza-Arenas}}]{Alipanah2025}%
  \BibitemOpen
  \bibfield  {author} {\bibinfo {author} {\bibfnamefont {H.}~\bibnamefont
  {Alipanah}}, \bibinfo {author} {\bibfnamefont {F.}~\bibnamefont {Zhang}},
  \bibinfo {author} {\bibfnamefont {Y.}~\bibnamefont {Yao}}, \bibinfo {author}
  {\bibfnamefont {R.}~\bibnamefont {Thompson}}, \bibinfo {author}
  {\bibfnamefont {N.}~\bibnamefont {Nguyen}}, \bibinfo {author} {\bibfnamefont
  {J.}~\bibnamefont {Liu}}, \bibinfo {author} {\bibfnamefont {P.}~\bibnamefont
  {Givi}}, \bibinfo {author} {\bibfnamefont {B.~J.}\ \bibnamefont
  {McDermott}},\ and\ \bibinfo {author} {\bibfnamefont {J.~J.}\ \bibnamefont
  {Mendoza-Arenas}},\ }\href {https://doi.org/10.1103/ndc3-bdwt} {\bibfield
  {journal} {\bibinfo  {journal} {Phys. Rev. Res.}\ ,\ } (\bibinfo {year}
  {2025})}\BibitemShut {NoStop}%
\bibitem [{\citenamefont {Brearley}\ and\ \citenamefont
  {Laizet}(2024)}]{Brearly2024}%
  \BibitemOpen
  \bibfield  {author} {\bibinfo {author} {\bibfnamefont {P.}~\bibnamefont
  {Brearley}}\ and\ \bibinfo {author} {\bibfnamefont {S.}~\bibnamefont
  {Laizet}},\ }\href {https://doi.org/10.1103/PhysRevA.110.012430} {\bibfield
  {journal} {\bibinfo  {journal} {Phys. Rev. A}\ }\textbf {\bibinfo {volume}
  {110}},\ \bibinfo {pages} {012430} (\bibinfo {year} {2024})}\BibitemShut
  {NoStop}%
\bibitem [{\citenamefont {Over}\ \emph
  {et~al.}(2025{\natexlab{b}})\citenamefont {Over}, \citenamefont {Bengoechea},
  \citenamefont {Brearley}, \citenamefont {Laizet},\ and\ \citenamefont
  {Rung}}]{Over2024b}%
  \BibitemOpen
  \bibfield  {author} {\bibinfo {author} {\bibfnamefont {P.}~\bibnamefont
  {Over}}, \bibinfo {author} {\bibfnamefont {S.}~\bibnamefont {Bengoechea}},
  \bibinfo {author} {\bibfnamefont {P.}~\bibnamefont {Brearley}}, \bibinfo
  {author} {\bibfnamefont {S.}~\bibnamefont {Laizet}},\ and\ \bibinfo {author}
  {\bibfnamefont {T.}~\bibnamefont {Rung}},\ }\href
  {https://doi.org/10.1103/d8hb-fv93} {\bibfield  {journal} {\bibinfo
  {journal} {Phys. Rev. A}\ }\textbf {\bibinfo {volume} {112}},\ \bibinfo
  {pages} {L010401} (\bibinfo {year} {2025}{\natexlab{b}})}\BibitemShut
  {NoStop}%
\bibitem [{\citenamefont {An}\ \emph {et~al.}(2023)\citenamefont {An},
  \citenamefont {Liu},\ and\ \citenamefont {Lin}}]{An2023}%
  \BibitemOpen
  \bibfield  {author} {\bibinfo {author} {\bibfnamefont {D.}~\bibnamefont
  {An}}, \bibinfo {author} {\bibfnamefont {J.-P.}\ \bibnamefont {Liu}},\ and\
  \bibinfo {author} {\bibfnamefont {L.}~\bibnamefont {Lin}},\ }\href
  {https://doi.org/10.1103/PhysRevLett.131.150603} {\bibfield  {journal}
  {\bibinfo  {journal} {Physical Review Letters}\ }\textbf {\bibinfo {volume}
  {131}},\ \bibinfo {pages} {150603} (\bibinfo {year} {2023})}\BibitemShut
  {NoStop}%
\bibitem [{\citenamefont {Costa}\ \emph {et~al.}(2019)\citenamefont {Costa},
  \citenamefont {Jordan},\ and\ \citenamefont {Ostrander}}]{Costa2019}%
  \BibitemOpen
  \bibfield  {author} {\bibinfo {author} {\bibfnamefont {P.~C.~S.}\
  \bibnamefont {Costa}}, \bibinfo {author} {\bibfnamefont {S.}~\bibnamefont
  {Jordan}},\ and\ \bibinfo {author} {\bibfnamefont {A.}~\bibnamefont
  {Ostrander}},\ }\href {https://doi.org/10.1103/PhysRevA.99.012323} {\bibfield
   {journal} {\bibinfo  {journal} {Phys. Rev. A}\ }\textbf {\bibinfo {volume}
  {99}},\ \bibinfo {pages} {012323} (\bibinfo {year} {2019})}\BibitemShut
  {NoStop}%
\bibitem [{\citenamefont {Esmaeilifar}\ \emph {et~al.}(2024)\citenamefont
  {Esmaeilifar}, \citenamefont {Ahn},\ and\ \citenamefont
  {Myong}}]{Esmaeilifar2024}%
  \BibitemOpen
  \bibfield  {author} {\bibinfo {author} {\bibfnamefont {E.}~\bibnamefont
  {Esmaeilifar}}, \bibinfo {author} {\bibfnamefont {D.}~\bibnamefont {Ahn}},\
  and\ \bibinfo {author} {\bibfnamefont {R.~S.}\ \bibnamefont {Myong}},\ }\href
  {https://doi.org/10.1063/5.0231994} {\bibfield  {journal} {\bibinfo
  {journal} {Physics of Fluids}\ }\textbf {\bibinfo {volume} {36}},\ \bibinfo
  {pages} {106110} (\bibinfo {year} {2024})}\BibitemShut {NoStop}%
\bibitem [{\citenamefont {Pool}\ \emph {et~al.}(2024)\citenamefont {Pool},
  \citenamefont {Somoza}, \citenamefont {Mc~Keever}, \citenamefont {Lubasch},\
  and\ \citenamefont {Horstmann}}]{Pool2024}%
  \BibitemOpen
  \bibfield  {author} {\bibinfo {author} {\bibfnamefont {A.~J.}\ \bibnamefont
  {Pool}}, \bibinfo {author} {\bibfnamefont {A.~D.}\ \bibnamefont {Somoza}},
  \bibinfo {author} {\bibfnamefont {C.}~\bibnamefont {Mc~Keever}}, \bibinfo
  {author} {\bibfnamefont {M.}~\bibnamefont {Lubasch}},\ and\ \bibinfo {author}
  {\bibfnamefont {B.}~\bibnamefont {Horstmann}},\ }\href
  {https://doi.org/10.1103/PhysRevResearch.6.033257} {\bibfield  {journal}
  {\bibinfo  {journal} {Phys. Rev. Res.}\ }\textbf {\bibinfo {volume} {6}},\
  \bibinfo {pages} {033257} (\bibinfo {year} {2024})}\BibitemShut {NoStop}%
\bibitem [{\citenamefont {Sanavio}\ \emph {et~al.}(2024)\citenamefont
  {Sanavio}, \citenamefont {Scatamacchia}, \citenamefont {de~Falco},\ and\
  \citenamefont {Succi}}]{Sanavio2024}%
  \BibitemOpen
  \bibfield  {author} {\bibinfo {author} {\bibfnamefont {C.}~\bibnamefont
  {Sanavio}}, \bibinfo {author} {\bibfnamefont {R.}~\bibnamefont
  {Scatamacchia}}, \bibinfo {author} {\bibfnamefont {C.}~\bibnamefont
  {de~Falco}},\ and\ \bibinfo {author} {\bibfnamefont {S.}~\bibnamefont
  {Succi}},\ }\href {https://doi.org/10.1063/5.0204955} {\bibfield  {journal}
  {\bibinfo  {journal} {Physics of Fluids}\ }\textbf {\bibinfo {volume} {36}},\
  \bibinfo {pages} {057143} (\bibinfo {year} {2024})}\BibitemShut {NoStop}%
\bibitem [{\citenamefont {Bharadwaj}\ and\ \citenamefont
  {Sreenivasan}(2025)}]{Bharadwaj2024}%
  \BibitemOpen
  \bibfield  {author} {\bibinfo {author} {\bibfnamefont {S.~S.}\ \bibnamefont
  {Bharadwaj}}\ and\ \bibinfo {author} {\bibfnamefont {K.~R.}\ \bibnamefont
  {Sreenivasan}},\ }\href {https://doi.org/10.1103/PhysRevResearch.7.023262}
  {\bibfield  {journal} {\bibinfo  {journal} {Phys. Rev. Res.}\ }\textbf
  {\bibinfo {volume} {7}},\ \bibinfo {pages} {023262} (\bibinfo {year}
  {2025})}\BibitemShut {NoStop}%
\bibitem [{\citenamefont {Fang}\ \emph {et~al.}(2023)\citenamefont {Fang},
  \citenamefont {Lin},\ and\ \citenamefont {Tong}}]{Fang2023}%
  \BibitemOpen
  \bibfield  {author} {\bibinfo {author} {\bibfnamefont {D.}~\bibnamefont
  {Fang}}, \bibinfo {author} {\bibfnamefont {L.}~\bibnamefont {Lin}},\ and\
  \bibinfo {author} {\bibfnamefont {Y.}~\bibnamefont {Tong}},\ }\href
  {https://doi.org/10.22331/q-2023-03-20-955} {\bibfield  {journal} {\bibinfo
  {journal} {Quantum}\ }\textbf {\bibinfo {volume} {7}},\ \bibinfo {pages}
  {955} (\bibinfo {year} {2023})}\BibitemShut {NoStop}%
\bibitem [{\citenamefont {Gily\'{e}n}\ \emph {et~al.}(2019)\citenamefont
  {Gily\'{e}n}, \citenamefont {Su}, \citenamefont {Low},\ and\ \citenamefont
  {Wiebe}}]{Gilyen2019}%
  \BibitemOpen
  \bibfield  {author} {\bibinfo {author} {\bibfnamefont {A.}~\bibnamefont
  {Gily\'{e}n}}, \bibinfo {author} {\bibfnamefont {Y.}~\bibnamefont {Su}},
  \bibinfo {author} {\bibfnamefont {G.~H.}\ \bibnamefont {Low}},\ and\ \bibinfo
  {author} {\bibfnamefont {N.}~\bibnamefont {Wiebe}},\ }in\ \href
  {https://doi.org/10.1145/3313276.3316366} {\emph {\bibinfo {booktitle}
  {Proceedings of the 51st Annual ACM SIGACT Symposium on Theory of
  Computing}}},\ \bibinfo {series and number} {STOC 2019}\ (\bibinfo
  {publisher} {Association for Computing Machinery},\ \bibinfo {address} {New
  York, NY, USA},\ \bibinfo {year} {2019})\ p.\ \bibinfo {pages}
  {193–204}\BibitemShut {NoStop}%
\bibitem [{\citenamefont {Camps}\ \emph {et~al.}(2024)\citenamefont {Camps},
  \citenamefont {Lin}, \citenamefont {Van~Beeumen},\ and\ \citenamefont
  {Yang}}]{Camps2024}%
  \BibitemOpen
  \bibfield  {author} {\bibinfo {author} {\bibfnamefont {D.}~\bibnamefont
  {Camps}}, \bibinfo {author} {\bibfnamefont {L.}~\bibnamefont {Lin}}, \bibinfo
  {author} {\bibfnamefont {R.}~\bibnamefont {Van~Beeumen}},\ and\ \bibinfo
  {author} {\bibfnamefont {C.}~\bibnamefont {Yang}},\ }\href
  {https://doi.org/10.1137/22M1484298} {\bibfield  {journal} {\bibinfo
  {journal} {SIAM Journal on Matrix Analysis and Applications}\ }\textbf
  {\bibinfo {volume} {45}},\ \bibinfo {pages} {801} (\bibinfo {year}
  {2024})}\BibitemShut {NoStop}%
\bibitem [{\citenamefont {Jin}\ \emph {et~al.}(2023)\citenamefont {Jin},
  \citenamefont {Liu},\ and\ \citenamefont {Yu}}]{Jin2023}%
  \BibitemOpen
  \bibfield  {author} {\bibinfo {author} {\bibfnamefont {S.}~\bibnamefont
  {Jin}}, \bibinfo {author} {\bibfnamefont {N.}~\bibnamefont {Liu}},\ and\
  \bibinfo {author} {\bibfnamefont {Y.}~\bibnamefont {Yu}},\ }\href
  {https://doi.org/10.1103/PhysRevA.108.032603} {\bibfield  {journal} {\bibinfo
   {journal} {Phys. Rev. A}\ }\textbf {\bibinfo {volume} {108}},\ \bibinfo
  {pages} {032603} (\bibinfo {year} {2023})}\BibitemShut {NoStop}%
\bibitem [{\citenamefont {Jin}\ \emph {et~al.}(2024{\natexlab{a}})\citenamefont
  {Jin}, \citenamefont {Li}, \citenamefont {Liu},\ and\ \citenamefont
  {Yu}}]{Jin2024}%
  \BibitemOpen
  \bibfield  {author} {\bibinfo {author} {\bibfnamefont {S.}~\bibnamefont
  {Jin}}, \bibinfo {author} {\bibfnamefont {X.}~\bibnamefont {Li}}, \bibinfo
  {author} {\bibfnamefont {N.}~\bibnamefont {Liu}},\ and\ \bibinfo {author}
  {\bibfnamefont {Y.}~\bibnamefont {Yu}},\ }\href
  {https://doi.org/https://doi.org/10.1016/j.jcp.2023.112707} {\bibfield
  {journal} {\bibinfo  {journal} {Journal of Computational Physics}\ }\textbf
  {\bibinfo {volume} {498}},\ \bibinfo {pages} {112707} (\bibinfo {year}
  {2024}{\natexlab{a}})}\BibitemShut {NoStop}%
\bibitem [{\citenamefont {Jin}\ \emph {et~al.}(2024{\natexlab{b}})\citenamefont
  {Jin}, \citenamefont {Li}, \citenamefont {Liu},\ and\ \citenamefont
  {Yu}}]{Jin2024b}%
  \BibitemOpen
  \bibfield  {author} {\bibinfo {author} {\bibfnamefont {S.}~\bibnamefont
  {Jin}}, \bibinfo {author} {\bibfnamefont {X.}~\bibnamefont {Li}}, \bibinfo
  {author} {\bibfnamefont {N.}~\bibnamefont {Liu}},\ and\ \bibinfo {author}
  {\bibfnamefont {Y.}~\bibnamefont {Yu}},\ }\href
  {https://doi.org/10.1137/23M1563451} {\bibfield  {journal} {\bibinfo
  {journal} {SIAM Journal on Scientific Computing}\ }\textbf {\bibinfo {volume}
  {46}},\ \bibinfo {pages} {B403} (\bibinfo {year}
  {2024}{\natexlab{b}})}\BibitemShut {NoStop}%
\bibitem [{\citenamefont {Jin}\ \emph {et~al.}(2024{\natexlab{c}})\citenamefont
  {Jin}, \citenamefont {Liu},\ and\ \citenamefont {Yu}}]{Jin2024c}%
  \BibitemOpen
  \bibfield  {author} {\bibinfo {author} {\bibfnamefont {S.}~\bibnamefont
  {Jin}}, \bibinfo {author} {\bibfnamefont {N.}~\bibnamefont {Liu}},\ and\
  \bibinfo {author} {\bibfnamefont {Y.}~\bibnamefont {Yu}},\ }\href
  {https://doi.org/10.1103/PhysRevLett.133.230602} {\bibfield  {journal}
  {\bibinfo  {journal} {Phys. Rev. Lett.}\ }\textbf {\bibinfo {volume} {133}},\
  \bibinfo {pages} {230602} (\bibinfo {year} {2024}{\natexlab{c}})}\BibitemShut
  {NoStop}%
\bibitem [{\citenamefont {Lu}\ and\ \citenamefont {Yang}(2024)}]{Lu2024}%
  \BibitemOpen
  \bibfield  {author} {\bibinfo {author} {\bibfnamefont {Z.}~\bibnamefont
  {Lu}}\ and\ \bibinfo {author} {\bibfnamefont {Y.}~\bibnamefont {Yang}},\
  }\href {https://doi.org/https://doi.org/10.1016/j.proci.2024.105440}
  {\bibfield  {journal} {\bibinfo  {journal} {Proceedings of the Combustion
  Institute}\ }\textbf {\bibinfo {volume} {40}},\ \bibinfo {pages} {105440}
  (\bibinfo {year} {2024})}\BibitemShut {NoStop}%
\bibitem [{\citenamefont {Childs}\ \emph {et~al.}(2021)\citenamefont {Childs},
  \citenamefont {Liu},\ and\ \citenamefont {Ostrander}}]{Childs2021}%
  \BibitemOpen
  \bibfield  {author} {\bibinfo {author} {\bibfnamefont {A.~M.}\ \bibnamefont
  {Childs}}, \bibinfo {author} {\bibfnamefont {J.-P.}\ \bibnamefont {Liu}},\
  and\ \bibinfo {author} {\bibfnamefont {A.}~\bibnamefont {Ostrander}},\ }\href
  {https://doi.org/10.22331/q-2021-11-10-574} {\bibfield  {journal} {\bibinfo
  {journal} {{Quantum}}\ }\textbf {\bibinfo {volume} {5}},\ \bibinfo {pages}
  {574} (\bibinfo {year} {2021})}\BibitemShut {NoStop}%
\bibitem [{\citenamefont {Griffiths}\ and\ \citenamefont
  {Schroeter}(2018)}]{Griffiths2018}%
  \BibitemOpen
  \bibfield  {author} {\bibinfo {author} {\bibfnamefont {D.~J.}\ \bibnamefont
  {Griffiths}}\ and\ \bibinfo {author} {\bibfnamefont {D.~F.}\ \bibnamefont
  {Schroeter}},\ }\href {https://doi.org/10.1017/9781316995433} {\emph
  {\bibinfo {title} {Introduction to Quantum Mechanics}}},\ \bibinfo {edition}
  {3rd}\ ed.\ (\bibinfo  {publisher} {Cambridge University Press},\ \bibinfo
  {year} {2018})\BibitemShut {NoStop}%
\bibitem [{\citenamefont {Ran}(2020)}]{Ran2020}%
  \BibitemOpen
  \bibfield  {author} {\bibinfo {author} {\bibfnamefont {S.-J.}\ \bibnamefont
  {Ran}},\ }\href {https://doi.org/10.1103/PhysRevA.101.032310} {\bibfield
  {journal} {\bibinfo  {journal} {Phys. Rev. A}\ }\textbf {\bibinfo {volume}
  {101}},\ \bibinfo {pages} {032310} (\bibinfo {year} {2020})}\BibitemShut
  {NoStop}%
\bibitem [{\citenamefont {Zhou}\ \emph {et~al.}(2021)\citenamefont {Zhou},
  \citenamefont {Hong},\ and\ \citenamefont {Ran}}]{Zhou2021}%
  \BibitemOpen
  \bibfield  {author} {\bibinfo {author} {\bibfnamefont {P.-F.}\ \bibnamefont
  {Zhou}}, \bibinfo {author} {\bibfnamefont {R.}~\bibnamefont {Hong}},\ and\
  \bibinfo {author} {\bibfnamefont {S.-J.}\ \bibnamefont {Ran}},\ }\href
  {https://doi.org/10.1103/PhysRevA.104.042601} {\bibfield  {journal} {\bibinfo
   {journal} {Phys. Rev. A}\ }\textbf {\bibinfo {volume} {104}},\ \bibinfo
  {pages} {042601} (\bibinfo {year} {2021})}\BibitemShut {NoStop}%
\bibitem [{\citenamefont {Termanova}\ \emph {et~al.}(2024)\citenamefont
  {Termanova}, \citenamefont {Melnikov}, \citenamefont {Mamenchikov},
  \citenamefont {Belokonev}, \citenamefont {Dolgov}, \citenamefont
  {Berezutskii}, \citenamefont {Ellerbrock}, \citenamefont {Mansell},\ and\
  \citenamefont {Perelshtein}}]{Termanova2024}%
  \BibitemOpen
  \bibfield  {author} {\bibinfo {author} {\bibfnamefont {A.}~\bibnamefont
  {Termanova}}, \bibinfo {author} {\bibfnamefont {A.}~\bibnamefont {Melnikov}},
  \bibinfo {author} {\bibfnamefont {E.}~\bibnamefont {Mamenchikov}}, \bibinfo
  {author} {\bibfnamefont {N.}~\bibnamefont {Belokonev}}, \bibinfo {author}
  {\bibfnamefont {S.}~\bibnamefont {Dolgov}}, \bibinfo {author} {\bibfnamefont
  {A.}~\bibnamefont {Berezutskii}}, \bibinfo {author} {\bibfnamefont
  {R.}~\bibnamefont {Ellerbrock}}, \bibinfo {author} {\bibfnamefont
  {C.}~\bibnamefont {Mansell}},\ and\ \bibinfo {author} {\bibfnamefont {M.~R.}\
  \bibnamefont {Perelshtein}},\ }\href
  {https://doi.org/10.1088/1367-2630/ad985b} {\bibfield  {journal} {\bibinfo
  {journal} {New Journal of Physics}\ }\textbf {\bibinfo {volume} {26}},\
  \bibinfo {pages} {123019} (\bibinfo {year} {2024})}\BibitemShut {NoStop}%
\bibitem [{\citenamefont {Oseledets}(2013)}]{Oseledets2013}%
  \BibitemOpen
  \bibfield  {author} {\bibinfo {author} {\bibfnamefont {I.~V.}\ \bibnamefont
  {Oseledets}},\ }\href {https://doi.org/10.1007/s00365-012-9175-x} {\bibfield
  {journal} {\bibinfo  {journal} {Constructive Approximation}\ }\textbf
  {\bibinfo {volume} {37}},\ \bibinfo {pages} {1} (\bibinfo {year}
  {2013})}\BibitemShut {NoStop}%
\bibitem [{\citenamefont {Berry}\ \emph
  {et~al.}(2015{\natexlab{a}})\citenamefont {Berry}, \citenamefont {Childs},\
  and\ \citenamefont {Kothari}}]{Berry2015}%
  \BibitemOpen
  \bibfield  {author} {\bibinfo {author} {\bibfnamefont {D.~W.}\ \bibnamefont
  {Berry}}, \bibinfo {author} {\bibfnamefont {A.~M.}\ \bibnamefont {Childs}},\
  and\ \bibinfo {author} {\bibfnamefont {R.}~\bibnamefont {Kothari}},\ }in\
  \href {https://doi.org/10.1109/FOCS.2015.54} {\emph {\bibinfo {booktitle}
  {2015 IEEE 56th Annual Symposium on Foundations of Computer Science}}}\
  (\bibinfo {year} {2015})\ pp.\ \bibinfo {pages} {792--809}\BibitemShut
  {NoStop}%
\bibitem [{\citenamefont {Childs}\ \emph {et~al.}(2018)\citenamefont {Childs},
  \citenamefont {Maslov}, \citenamefont {Nam}, \citenamefont {Ross},\ and\
  \citenamefont {Su}}]{Childs2018}%
  \BibitemOpen
  \bibfield  {author} {\bibinfo {author} {\bibfnamefont {A.~M.}\ \bibnamefont
  {Childs}}, \bibinfo {author} {\bibfnamefont {D.}~\bibnamefont {Maslov}},
  \bibinfo {author} {\bibfnamefont {Y.}~\bibnamefont {Nam}}, \bibinfo {author}
  {\bibfnamefont {N.~J.}\ \bibnamefont {Ross}},\ and\ \bibinfo {author}
  {\bibfnamefont {Y.}~\bibnamefont {Su}},\ }\href
  {https://doi.org/10.1073/pnas.1801723115} {\bibfield  {journal} {\bibinfo
  {journal} {Proceedings of the National Academy of Sciences}\ }\textbf
  {\bibinfo {volume} {115}},\ \bibinfo {pages} {9456} (\bibinfo {year}
  {2018})}\BibitemShut {NoStop}%
\bibitem [{\citenamefont {Hatano}\ and\ \citenamefont
  {Suzuki}(2005)}]{Hatano2005}%
  \BibitemOpen
  \bibfield  {author} {\bibinfo {author} {\bibfnamefont {N.}~\bibnamefont
  {Hatano}}\ and\ \bibinfo {author} {\bibfnamefont {M.}~\bibnamefont
  {Suzuki}},\ }\bibinfo {title} {Finding exponential product formulas of higher
  orders},\ in\ \href {https://doi.org/10.1007/11526216_2} {\emph {\bibinfo
  {booktitle} {"Quantum Annealing and Other Optimization Methods}}}\ (\bibinfo
  {publisher} {Springer Berlin Heidelberg},\ \bibinfo {year} {2005})\ pp.\
  \bibinfo {pages} {37--68}\BibitemShut {NoStop}%
\bibitem [{\citenamefont {Berry}\ \emph
  {et~al.}(2015{\natexlab{b}})\citenamefont {Berry}, \citenamefont {Childs},
  \citenamefont {Cleve}, \citenamefont {Kothari},\ and\ \citenamefont
  {Somma}}]{Berry2015_Taylor}%
  \BibitemOpen
  \bibfield  {author} {\bibinfo {author} {\bibfnamefont {D.~W.}\ \bibnamefont
  {Berry}}, \bibinfo {author} {\bibfnamefont {A.~M.}\ \bibnamefont {Childs}},
  \bibinfo {author} {\bibfnamefont {R.}~\bibnamefont {Cleve}}, \bibinfo
  {author} {\bibfnamefont {R.}~\bibnamefont {Kothari}},\ and\ \bibinfo {author}
  {\bibfnamefont {R.~D.}\ \bibnamefont {Somma}},\ }\href
  {https://doi.org/10.1103/PhysRevLett.114.090502} {\bibfield  {journal}
  {\bibinfo  {journal} {Phys. Rev. Lett.}\ }\textbf {\bibinfo {volume} {114}},\
  \bibinfo {pages} {090502} (\bibinfo {year} {2015}{\natexlab{b}})}\BibitemShut
  {NoStop}%
\bibitem [{\citenamefont {Low}\ and\ \citenamefont {Chuang}(2017)}]{Low2017}%
  \BibitemOpen
  \bibfield  {author} {\bibinfo {author} {\bibfnamefont {G.~H.}\ \bibnamefont
  {Low}}\ and\ \bibinfo {author} {\bibfnamefont {I.~L.}\ \bibnamefont
  {Chuang}},\ }\href {https://doi.org/10.1103/PhysRevLett.118.010501}
  {\bibfield  {journal} {\bibinfo  {journal} {Phys. Rev. Lett.}\ }\textbf
  {\bibinfo {volume} {118}},\ \bibinfo {pages} {010501} (\bibinfo {year}
  {2017})}\BibitemShut {NoStop}%
\bibitem [{\citenamefont {Gingrich}\ and\ \citenamefont
  {Williams}(2004)}]{Gingrich2004}%
  \BibitemOpen
  \bibfield  {author} {\bibinfo {author} {\bibfnamefont {R.~M.}\ \bibnamefont
  {Gingrich}}\ and\ \bibinfo {author} {\bibfnamefont {C.~P.}\ \bibnamefont
  {Williams}},\ }in\ \href {https://dl.acm.org/doi/10.5555/984720.984757}
  {\emph {\bibinfo {booktitle} {Proceedings of the Winter International
  Synposium on Information and Communication Technologies}}},\ Vol.~\bibinfo
  {volume} {58}\ (\bibinfo {year} {2004})\ pp.\ \bibinfo {pages}
  {1--6}\BibitemShut {NoStop}%
\bibitem [{\citenamefont {Lapworth}\ and\ \citenamefont
  {Sünderhauf}(2025)}]{LapworthSuenderhauf2025}%
  \BibitemOpen
  \bibfield  {author} {\bibinfo {author} {\bibfnamefont {L.}~\bibnamefont
  {Lapworth}}\ and\ \bibinfo {author} {\bibfnamefont {C.}~\bibnamefont
  {Sünderhauf}},\ }\href {https://doi.org/10.1088/2058-9565/ae0f4b} {\bibfield
   {journal} {\bibinfo  {journal} {Quantum Science and Technology}\ }\textbf
  {\bibinfo {volume} {10}},\ \bibinfo {pages} {045064} (\bibinfo {year}
  {2025})}\BibitemShut {NoStop}%
\bibitem [{\citenamefont {Lin}\ \emph {et~al.}(2021)\citenamefont {Lin},
  \citenamefont {Dilip}, \citenamefont {Green}, \citenamefont {Smith},\ and\
  \citenamefont {Pollmann}}]{Lin2021}%
  \BibitemOpen
  \bibfield  {author} {\bibinfo {author} {\bibfnamefont {S.-H.}\ \bibnamefont
  {Lin}}, \bibinfo {author} {\bibfnamefont {R.}~\bibnamefont {Dilip}}, \bibinfo
  {author} {\bibfnamefont {A.~G.}\ \bibnamefont {Green}}, \bibinfo {author}
  {\bibfnamefont {A.}~\bibnamefont {Smith}},\ and\ \bibinfo {author}
  {\bibfnamefont {F.}~\bibnamefont {Pollmann}},\ }\href
  {https://doi.org/10.1103/PRXQuantum.2.010342} {\bibfield  {journal} {\bibinfo
   {journal} {PRX Quantum}\ }\textbf {\bibinfo {volume} {2}},\ \bibinfo {pages}
  {010342} (\bibinfo {year} {2021})}\BibitemShut {NoStop}%
\bibitem [{\citenamefont {Childs}\ and\ \citenamefont
  {Wiebe}(2012)}]{Childs2012}%
  \BibitemOpen
  \bibfield  {author} {\bibinfo {author} {\bibfnamefont {A.~M.}\ \bibnamefont
  {Childs}}\ and\ \bibinfo {author} {\bibfnamefont {N.}~\bibnamefont {Wiebe}},\
  }\href {https://doi.org/10.26421/QIC12.11-12-1} {\bibfield  {journal}
  {\bibinfo  {journal} {Quantum Information \& Computation}\ }\textbf {\bibinfo
  {volume} {12}},\ \bibinfo {pages} {901} (\bibinfo {year} {2012})}\BibitemShut
  {NoStop}%
\bibitem [{\citenamefont {Hantzko}\ \emph {et~al.}(2024)\citenamefont
  {Hantzko}, \citenamefont {Binkowski},\ and\ \citenamefont
  {Gupta}}]{Hantzko2024}%
  \BibitemOpen
  \bibfield  {author} {\bibinfo {author} {\bibfnamefont {L.}~\bibnamefont
  {Hantzko}}, \bibinfo {author} {\bibfnamefont {L.}~\bibnamefont {Binkowski}},\
  and\ \bibinfo {author} {\bibfnamefont {S.}~\bibnamefont {Gupta}},\ }\href
  {https://doi.org/10.1088/1402-4896/ad6499} {\bibfield  {journal} {\bibinfo
  {journal} {Physica Scripta}\ }\textbf {\bibinfo {volume} {99}},\ \bibinfo
  {pages} {085128} (\bibinfo {year} {2024})}\BibitemShut {NoStop}%
\bibitem [{\citenamefont {Sano}\ and\ \citenamefont
  {Hamamura}(2025)}]{Sano2024}%
  \BibitemOpen
  \bibfield  {author} {\bibinfo {author} {\bibfnamefont {Y.}~\bibnamefont
  {Sano}}\ and\ \bibinfo {author} {\bibfnamefont {I.}~\bibnamefont
  {Hamamura}},\ }\href {https://doi.org/10.1103/yqj3-hyxv} {\bibfield
  {journal} {\bibinfo  {journal} {Phys. Rev. Appl.}\ }\textbf {\bibinfo
  {volume} {24}},\ \bibinfo {pages} {034062} (\bibinfo {year}
  {2025})}\BibitemShut {NoStop}%
\bibitem [{\citenamefont {van Hülst}\ \emph {et~al.}(2025)\citenamefont {van
  Hülst}, \citenamefont {Siegl}, \citenamefont {Over}, \citenamefont
  {Bengoechea}, \citenamefont {Hashizume}, \citenamefont {Cecile},
  \citenamefont {Rung},\ and\ \citenamefont {Jaksch}}]{vanHuelst2025}%
  \BibitemOpen
  \bibfield  {author} {\bibinfo {author} {\bibfnamefont {N.-L.}\ \bibnamefont
  {van Hülst}}, \bibinfo {author} {\bibfnamefont {P.}~\bibnamefont {Siegl}},
  \bibinfo {author} {\bibfnamefont {P.}~\bibnamefont {Over}}, \bibinfo {author}
  {\bibfnamefont {S.}~\bibnamefont {Bengoechea}}, \bibinfo {author}
  {\bibfnamefont {T.}~\bibnamefont {Hashizume}}, \bibinfo {author}
  {\bibfnamefont {M.~G.}\ \bibnamefont {Cecile}}, \bibinfo {author}
  {\bibfnamefont {T.}~\bibnamefont {Rung}},\ and\ \bibinfo {author}
  {\bibfnamefont {D.}~\bibnamefont {Jaksch}},\ }\href
  {https://arxiv.org/abs/2507.05222} {\bibinfo {title} {Quantum-inspired
  tensor-network fractional-step method for incompressible flow in curvilinear
  coordinates}} (\bibinfo {year} {2025}),\ \Eprint
  {https://arxiv.org/abs/2507.05222} {arXiv:2507.05222} \BibitemShut {NoStop}%
\bibitem [{\citenamefont {Siegl}\ \emph {et~al.}(2025)\citenamefont {Siegl},
  \citenamefont {Reese}, \citenamefont {Hashizume}, \citenamefont {van
  Hülst},\ and\ \citenamefont {Jaksch}}]{Siegl2025}%
  \BibitemOpen
  \bibfield  {author} {\bibinfo {author} {\bibfnamefont {P.}~\bibnamefont
  {Siegl}}, \bibinfo {author} {\bibfnamefont {G.~S.}\ \bibnamefont {Reese}},
  \bibinfo {author} {\bibfnamefont {T.}~\bibnamefont {Hashizume}}, \bibinfo
  {author} {\bibfnamefont {N.-L.}\ \bibnamefont {van Hülst}},\ and\ \bibinfo
  {author} {\bibfnamefont {D.}~\bibnamefont {Jaksch}},\ }\href
  {https://arxiv.org/abs/2502.04425} {\bibinfo {title} {Tensor-programmable
  quantum circuits for solving differential equations}} (\bibinfo {year}
  {2025}),\ \Eprint {https://arxiv.org/abs/2502.04425} {arXiv:2502.04425}
  \BibitemShut {NoStop}%
\end{thebibliography}%

\appendix
\section{LCU circuit to matrix}
\label{sec:appendix}
As an example of state-vector simulation, the matrix-vector operations executed by the LCU circuit of Fig.~\ref{fig:lcu:onedim} are illustrated here. 
Starting from the initial state 
\begin{equation}
    \ket{0 0 \, \phi^t}=\ket{0}\otimes\ket{0}\otimes\ket{\phi^t} \in \mathbb{R}^{2^{n+n_{\text{anc.}}}} \, , 
    \label{eq:LCU_init_state}
\end{equation}
the circuit sequentially applies the quantum gates $V_k,U_0, U_1, U_2, U_3, V_k^{\dagger}$, each represented by the corresponding matrix $\tilde{V}_k, \tilde{U}_0, \tilde{U}_1, \tilde{U}_2, \tilde{U}_3, \tilde{V}_k^{\dagger} \in \mathbb{R}^{2^{n+n_{\text{anc.}}} \times 2^{n+n_{\text{anc.}}}}$, evolving into the final state 
\begin{equation}
    \begin{split}
    \ket{00}\ket{\phi^{t+1}} + & \ket{01}\ket{*} + \ket{10}\ket{*} + \ket{11}\ket{*} = \\ 
    & \tilde{V}_k^{\dagger} \tilde{U}_3 \tilde{U}_2 \tilde{U}_1 \tilde{U}_0 \tilde{V}_k \ket{00\phi^t}\, . 
    \label{eq:LCU_matrix_vector}
    \end{split}
\end{equation}
The matrix $\tilde{V}_k$ is built with the matrices $V_k \in \mathbb{R}^{2^{n_{\text{anc.}}} \times 2^{n_{\text{anc.}}}}$ and $I \in \mathbb{R}^{2^{n} \times 2^{n}}$ as 
\begin{equation}
    \tilde{V}_k = V_k \otimes I = 
    \begin{bmatrix}
        V_{k_{00}}I & V_{k_{01}}I & V_{k_{02}}I & V_{k_{03}}I \\
        V_{k_{10}}I & V_{k_{11}}I & V_{k_{12}}I & V_{k_{13}}I \\
        V_{k_{20}}I & V_{k_{21}}I & V_{k_{22}}I & V_{k_{23}}I \\
        V_{k_{30}}I & V_{k_{31}}I & V_{k_{32}}I & V_{k_{33}}I
    \end{bmatrix}\, ,
    \label{eq:Vk_matrix}
\end{equation}
while the $00$-controlled gate $U_0$, the $01$-controlled gate $U_1$, the $10$-controlled gate $U_2$, and the $11$-controlled gate $U_3$ result in the matrices
\begin{equation}
\begin{split}
\tilde{U}_0 =  
    \begin{bmatrix}
        U_0 &   &   &   \\
            & I &   &   \\
            &   & I &   \\
            &   &   & I
    \end{bmatrix}, \, 
\tilde{U}_1 =  
    \begin{bmatrix}
        I &     &   &   \\
          & U_1 &   &   \\
          &     & I &   \\
          &     &   & I
    \end{bmatrix}, \, \\
\tilde{U}_2 =  
    \begin{bmatrix}
        I &   &    &   \\
          & I &    &   \\
          &   & U_2 &   \\
          &   &    & I
    \end{bmatrix}, \, \text{ and } 
\tilde{U}_3 = 
    \begin{bmatrix}
        I &   &   &   \\
          & I &   &   \\
          &   & I &   \\
          &   &   & U_3
    \end{bmatrix} \, .
\end{split}
\end{equation}

\end{document}